\documentclass[conference]{IEEEtran}
\usepackage{amsmath}           
\usepackage{amsfonts}   
\usepackage{amssymb}
\usepackage{subfigure}
\usepackage{graphicx}
\usepackage{algorithm}
\usepackage{algpseudocode}
\usepackage{algorithmicx}

\ifCLASSINFOpdf
\else
\fi
\newcommand{\nmotSizeI}{1.8ex}
\newcommand{\nmotSizeII}{2.8ex}
\newcommand{\nmotSizeIII}{1.8ex}
\newcommand{\nmotSizeIV}{0.9ex}
\newcommand{\motSizeI}{1.8ex}
\newcommand{\motSizeII}{2.8ex}
\newcommand{\motSizeIII}{1.8ex}
\newcommand{\motSizeIV}{0.9ex}
\newcommand{\nmotI}{
 {\mathchoice
  {\includegraphics[height=\nmotSizeI]{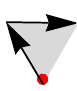}}
  {\includegraphics[height=\nmotSizeII]{nsp_triadConfig_1.pdf}}
  {\includegraphics[height=\nmotSizeIII]{nsp_triadConfig_1.pdf}}
  {\includegraphics[height=\nmotSizeIV]{nsp_triadConfig_1.pdf}}
 }
}
\newcommand{\nmotII}{
 {\mathchoice
  {\includegraphics[height=\nmotSizeI]{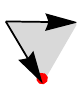}}
  {\includegraphics[height=\nmotSizeII]{nsp_triadConfig_2.pdf}}
  {\includegraphics[height=\nmotSizeIII]{nsp_triadConfig_2.pdf}}
  {\includegraphics[height=\nmotSizeIV]{nsp_triadConfig_2.pdf}}
 }
}
\newcommand{\nmotIII}{
 {\mathchoice
  {\includegraphics[height=\nmotSizeI]{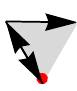}}
  {\includegraphics[height=\nmotSizeII]{nsp_triadConfig_3.pdf}}
  {\includegraphics[height=\nmotSizeIII]{nsp_triadConfig_3.pdf}}
  {\includegraphics[height=\nmotSizeIV]{nsp_triadConfig_3.pdf}}
 }
}
\newcommand{\nmotIV}{
 {\mathchoice
  {\includegraphics[height=\nmotSizeI]{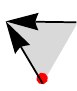}}
  {\includegraphics[height=\nmotSizeII]{nsp_triadConfig_4.pdf}}
  {\includegraphics[height=\nmotSizeIII]{nsp_triadConfig_4.pdf}}
  {\includegraphics[height=\nmotSizeIV]{nsp_triadConfig_4.pdf}}
 }
}
\newcommand{\nmotV}{
 {\mathchoice
  {\includegraphics[height=\nmotSizeI]{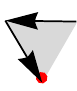}}
  {\includegraphics[height=\nmotSizeII]{nsp_triadConfig_5.pdf}}
  {\includegraphics[height=\nmotSizeIII]{nsp_triadConfig_5.pdf}}
  {\includegraphics[height=\nmotSizeIV]{nsp_triadConfig_5.pdf}}
 }
}
\newcommand{\nmotVI}{
 {\mathchoice
  {\includegraphics[height=\nmotSizeI]{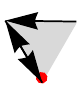}}
  {\includegraphics[height=\nmotSizeII]{nsp_triadConfig_6.pdf}}
  {\includegraphics[height=\nmotSizeIII]{nsp_triadConfig_6.pdf}}
  {\includegraphics[height=\nmotSizeIV]{nsp_triadConfig_6.pdf}}
 }
}
\newcommand{\nmotVII}{
 {\mathchoice
  {\includegraphics[height=\nmotSizeI]{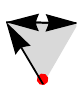}}
  {\includegraphics[height=\nmotSizeII]{nsp_triadConfig_7.pdf}}
  {\includegraphics[height=\nmotSizeIII]{nsp_triadConfig_7.pdf}}
  {\includegraphics[height=\nmotSizeIV]{nsp_triadConfig_7.pdf}}
 }
}
\newcommand{\nmotVIII}{
 {\mathchoice
  {\includegraphics[height=\nmotSizeI]{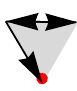}}
  {\includegraphics[height=\nmotSizeII]{nsp_triadConfig_8.pdf}}
  {\includegraphics[height=\nmotSizeIII]{nsp_triadConfig_8.pdf}}
  {\includegraphics[height=\nmotSizeIV]{nsp_triadConfig_8.pdf}}
 }
}
\newcommand{\nmotIX}{
 {\mathchoice
  {\includegraphics[height=\nmotSizeI]{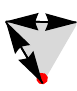}}
  {\includegraphics[height=\nmotSizeII]{nsp_triadConfig_9.pdf}}
  {\includegraphics[height=\nmotSizeIII]{nsp_triadConfig_9.pdf}}
  {\includegraphics[height=\nmotSizeIV]{nsp_triadConfig_9.pdf}}
 }
}
\newcommand{\nmotX}{
 {\mathchoice
  {\includegraphics[height=\nmotSizeI]{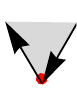}}
  {\includegraphics[height=\nmotSizeII]{nsp_triadConfig_10.pdf}}
  {\includegraphics[height=\nmotSizeIII]{nsp_triadConfig_10.pdf}}
  {\includegraphics[height=\nmotSizeIV]{nsp_triadConfig_10.pdf}}
 }
}
\newcommand{\nmotXI}{
 {\mathchoice
  {\includegraphics[height=\nmotSizeI]{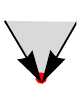}}
  {\includegraphics[height=\nmotSizeII]{nsp_triadConfig_11.pdf}}
  {\includegraphics[height=\nmotSizeIII]{nsp_triadConfig_11.pdf}}
  {\includegraphics[height=\nmotSizeIV]{nsp_triadConfig_11.pdf}}
 }
}
\newcommand{\nmotXII}{
 {\mathchoice
  {\includegraphics[height=\nmotSizeI]{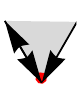}}
  {\includegraphics[height=\nmotSizeII]{nsp_triadConfig_12.pdf}}
  {\includegraphics[height=\nmotSizeIII]{nsp_triadConfig_12.pdf}}
  {\includegraphics[height=\nmotSizeIV]{nsp_triadConfig_12.pdf}}
 }
}
\newcommand{\nmotXIII}{
 {\mathchoice
  {\includegraphics[height=\nmotSizeI]{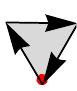}}
  {\includegraphics[height=\nmotSizeII]{nsp_triadConfig_13.pdf}}
  {\includegraphics[height=\nmotSizeIII]{nsp_triadConfig_13.pdf}}
  {\includegraphics[height=\nmotSizeIV]{nsp_triadConfig_13.pdf}}
 }
}
\newcommand{\nmotXIV}{
 {\mathchoice
  {\includegraphics[height=\nmotSizeI]{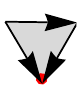}}
  {\includegraphics[height=\nmotSizeII]{nsp_triadConfig_14.pdf}}
  {\includegraphics[height=\nmotSizeIII]{nsp_triadConfig_14.pdf}}
  {\includegraphics[height=\nmotSizeIV]{nsp_triadConfig_14.pdf}}
 }
}
\newcommand{\nmotXV}{
 {\mathchoice
  {\includegraphics[height=\nmotSizeI]{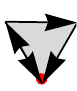}}
  {\includegraphics[height=\nmotSizeII]{nsp_triadConfig_15.pdf}}
  {\includegraphics[height=\nmotSizeIII]{nsp_triadConfig_15.pdf}}
  {\includegraphics[height=\nmotSizeIV]{nsp_triadConfig_15.pdf}}
 }
}
\newcommand{\nmotXVI}{
 {\mathchoice
  {\includegraphics[height=\nmotSizeI]{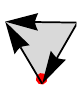}}
  {\includegraphics[height=\nmotSizeII]{nsp_triadConfig_16.pdf}}
  {\includegraphics[height=\nmotSizeIII]{nsp_triadConfig_16.pdf}}
  {\includegraphics[height=\nmotSizeIV]{nsp_triadConfig_16.pdf}}
 }
}
\newcommand{\nmotXVII}{
 {\mathchoice
  {\includegraphics[height=\nmotSizeI]{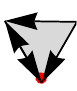}}
  {\includegraphics[height=\nmotSizeII]{nsp_triadConfig_17.pdf}}
  {\includegraphics[height=\nmotSizeIII]{nsp_triadConfig_17.pdf}}
  {\includegraphics[height=\nmotSizeIV]{nsp_triadConfig_17.pdf}}
 }
}
\newcommand{\nmotXVIII}{
 {\mathchoice
  {\includegraphics[height=\nmotSizeI]{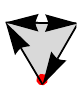}}
  {\includegraphics[height=\nmotSizeII]{nsp_triadConfig_18.pdf}}
  {\includegraphics[height=\nmotSizeIII]{nsp_triadConfig_18.pdf}}
  {\includegraphics[height=\nmotSizeIV]{nsp_triadConfig_18.pdf}}
 }
}
\newcommand{\nmotXIX}{
 {\mathchoice
  {\includegraphics[height=\nmotSizeI]{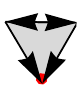}}
  {\includegraphics[height=\nmotSizeII]{nsp_triadConfig_19.pdf}}
  {\includegraphics[height=\nmotSizeIII]{nsp_triadConfig_19.pdf}}
  {\includegraphics[height=\nmotSizeIV]{nsp_triadConfig_19.pdf}}
 }
}
\newcommand{\nmotXX}{
 {\mathchoice
  {\includegraphics[height=\nmotSizeI]{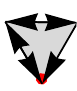}}
  {\includegraphics[height=\nmotSizeII]{nsp_triadConfig_20.pdf}}
  {\includegraphics[height=\nmotSizeIII]{nsp_triadConfig_20.pdf}}
  {\includegraphics[height=\nmotSizeIV]{nsp_triadConfig_20.pdf}}
 }
}
\newcommand{\nmotXXI}{
 {\mathchoice
  {\includegraphics[height=\nmotSizeI]{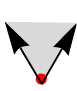}}
  {\includegraphics[height=\nmotSizeII]{nsp_triadConfig_21.pdf}}
  {\includegraphics[height=\nmotSizeIII]{nsp_triadConfig_21.pdf}}
  {\includegraphics[height=\nmotSizeIV]{nsp_triadConfig_21.pdf}}
 }
}
\newcommand{\nmotXXII}{
 {\mathchoice
  {\includegraphics[height=\nmotSizeI]{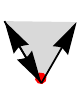}}
  {\includegraphics[height=\nmotSizeII]{nsp_triadConfig_22.pdf}}
  {\includegraphics[height=\nmotSizeIII]{nsp_triadConfig_22.pdf}}
  {\includegraphics[height=\nmotSizeIV]{nsp_triadConfig_22.pdf}}
 }
}
\newcommand{\nmotXXIII}{
 {\mathchoice
  {\includegraphics[height=\nmotSizeI]{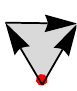}}
  {\includegraphics[height=\nmotSizeII]{nsp_triadConfig_23.pdf}}
  {\includegraphics[height=\nmotSizeIII]{nsp_triadConfig_23.pdf}}
  {\includegraphics[height=\nmotSizeIV]{nsp_triadConfig_23.pdf}}
 }
}

\newcommand{\nmotXXV}{
 {\mathchoice
  {\includegraphics[height=\nmotSizeI]{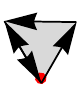}}
  {\includegraphics[height=\nmotSizeII]{nsp_triadConfig_25.pdf}}
  {\includegraphics[height=\nmotSizeIII]{nsp_triadConfig_25.pdf}}
  {\includegraphics[height=\nmotSizeIV]{nsp_triadConfig_25.pdf}}
 }
}
\newcommand{\nmotXXVI}{
 {\mathchoice
  {\includegraphics[height=\nmotSizeI]{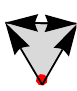}}
  {\includegraphics[height=\nmotSizeII]{nsp_triadConfig_26.pdf}}
  {\includegraphics[height=\nmotSizeIII]{nsp_triadConfig_26.pdf}}
  {\includegraphics[height=\nmotSizeIV]{nsp_triadConfig_26.pdf}}
 }
}
\newcommand{\nmotXXVII}{
 {\mathchoice
  {\includegraphics[height=\nmotSizeI]{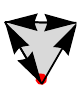}}
  {\includegraphics[height=\nmotSizeII]{nsp_triadConfig_27.pdf}}
  {\includegraphics[height=\nmotSizeIII]{nsp_triadConfig_27.pdf}}
  {\includegraphics[height=\nmotSizeIV]{nsp_triadConfig_27.pdf}}
 }
}

\newcommand{\nmotXXIX}{
 {\mathchoice
  {\includegraphics[height=\nmotSizeI]{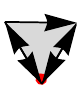}}
  {\includegraphics[height=\nmotSizeII]{nsp_triadConfig_29.pdf}}
  {\includegraphics[height=\nmotSizeIII]{nsp_triadConfig_29.pdf}}
  {\includegraphics[height=\nmotSizeIV]{nsp_triadConfig_29.pdf}}
 }
}
\newcommand{\nmotXXX}{
 {\mathchoice
  {\includegraphics[height=\nmotSizeI]{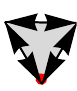}}
  {\includegraphics[height=\nmotSizeII]{nsp_triadConfig_30.pdf}}
  {\includegraphics[height=\nmotSizeIII]{nsp_triadConfig_30.pdf}}
  {\includegraphics[height=\nmotSizeIV]{nsp_triadConfig_30.pdf}}
 }
}

\newcommand{\motIV}{
 {\mathchoice
  {\includegraphics[height=\motSizeI]{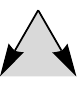}}
  {\includegraphics[height=\motSizeII]{triadConfig_4.pdf}}
  {\includegraphics[height=\motSizeIII]{triadConfig_4.pdf}}
  {\includegraphics[height=\motSizeIV]{triadConfig_4.pdf}}
 }
}
\newcommand{\motV}{
 {\mathchoice
  {\includegraphics[height=\motSizeI]{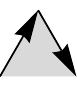}}
  {\includegraphics[height=\motSizeII]{triadConfig_5.pdf}}
  {\includegraphics[height=\motSizeIII]{triadConfig_5.pdf}}
  {\includegraphics[height=\motSizeIV]{triadConfig_5.pdf}}
 }
}
\newcommand{\motVI}{
 {\mathchoice
  {\includegraphics[height=\motSizeI]{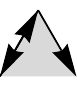}}
  {\includegraphics[height=\motSizeII]{triadConfig_6.pdf}}
  {\includegraphics[height=\motSizeIII]{triadConfig_6.pdf}}
  {\includegraphics[height=\motSizeIV]{triadConfig_6.pdf}}
 }
}
\newcommand{\motVII}{
 {\mathchoice
  {\includegraphics[height=\motSizeI]{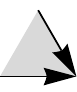}}
  {\includegraphics[height=\motSizeII]{triadConfig_7.pdf}}
  {\includegraphics[height=\motSizeIII]{triadConfig_7.pdf}}
  {\includegraphics[height=\motSizeIV]{triadConfig_7.pdf}}
 }
}
\newcommand{\motVIII}{
 {\mathchoice
  {\includegraphics[height=\motSizeI]{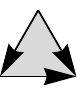}}
  {\includegraphics[height=\motSizeII]{triadConfig_8.pdf}}
  {\includegraphics[height=\motSizeIII]{triadConfig_8.pdf}}
  {\includegraphics[height=\motSizeIV]{triadConfig_8.pdf}}
 }
}
\newcommand{\motIX}{
 {\mathchoice
  {\includegraphics[height=\motSizeI]{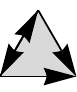}}
  {\includegraphics[height=\motSizeII]{triadConfig_9.pdf}}
  {\includegraphics[height=\motSizeIII]{triadConfig_9.pdf}}
  {\includegraphics[height=\motSizeIV]{triadConfig_9.pdf}}
 }
}
\newcommand{\motX}{
 {\mathchoice
  {\includegraphics[height=\motSizeI]{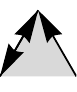}}
  {\includegraphics[height=\motSizeII]{triadConfig_10.pdf}}
  {\includegraphics[height=\motSizeIII]{triadConfig_10.pdf}}
  {\includegraphics[height=\motSizeIV]{triadConfig_10.pdf}}
 }
}
\newcommand{\motXI}{
 {\mathchoice
  {\includegraphics[height=\motSizeI]{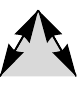}}
  {\includegraphics[height=\motSizeII]{triadConfig_11.pdf}}
  {\includegraphics[height=\motSizeIII]{triadConfig_11.pdf}}
  {\includegraphics[height=\motSizeIV]{triadConfig_11.pdf}}
 }
}
\newcommand{\motXII}{
 {\mathchoice
  {\includegraphics[height=\motSizeI]{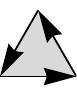}}
  {\includegraphics[height=\motSizeII]{triadConfig_12.pdf}}
  {\includegraphics[height=\motSizeIII]{triadConfig_12.pdf}}
  {\includegraphics[height=\motSizeIV]{triadConfig_12.pdf}}
 }
}
\newcommand{\motXIII}{
 {\mathchoice
  {\includegraphics[height=\motSizeI]{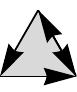}}
  {\includegraphics[height=\motSizeII]{triadConfig_13.pdf}}
  {\includegraphics[height=\motSizeIII]{triadConfig_13.pdf}}
  {\includegraphics[height=\motSizeIV]{triadConfig_13.pdf}}
 }
}
\newcommand{\motXIV}{
 {\mathchoice
  {\includegraphics[height=\motSizeI]{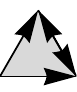}}
  {\includegraphics[height=\motSizeII]{triadConfig_14.pdf}}
  {\includegraphics[height=\motSizeIII]{triadConfig_14.pdf}}
  {\includegraphics[height=\motSizeIV]{triadConfig_14.pdf}}
 }
}
\newcommand{\motXV}{
 {\mathchoice
  {\includegraphics[height=\motSizeI]{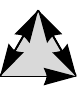}}
  {\includegraphics[height=\motSizeII]{triadConfig_15.pdf}}
  {\includegraphics[height=\motSizeIII]{triadConfig_15.pdf}}
  {\includegraphics[height=\motSizeIV]{triadConfig_15.pdf}}
 }
}
\newcommand{\motXVI}{
 {\mathchoice
  {\includegraphics[height=\motSizeI]{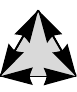}}
  {\includegraphics[height=\motSizeII]{triadConfig_16.pdf}}
  {\includegraphics[height=\motSizeIII]{triadConfig_16.pdf}}
  {\includegraphics[height=\motSizeIV]{triadConfig_16.pdf}}
 }
}

\begin{document}
%
\title{Node-Specific Triad Pattern Mining for Complex-Network Analysis}


\author{\IEEEauthorblockN{Marco Winkler}
\IEEEauthorblockA{Institute for Theoretical Physics\\University of Wuerzburg\\
Am Hubland, 97074 Wuerzburg, Germany\\
Email: mwinkler@physik.uni-wuerzburg.de}
\and
\IEEEauthorblockN{Joerg Reichardt}
\IEEEauthorblockA{Institute for Theoretical Physics\\University of Wuerzburg\\
Am Hubland, 97074 Wuerzburg, Germany\\
Email: reichardt@physik.uni-wuerzburg.de}}


\newcommand\blfootnote[1]{%
  \begingroup
  \renewcommand\thefootnote{}\footnote{#1}%
  \addtocounter{footnote}{-1}%
  \endgroup
}

\maketitle

\begin{abstract}
The mining of graphs in terms of their local substructure is a well-established methodology to analyze networks. It was hypothesized that motifs - subgraph patterns which appear significantly more often than expected at random - play a key role for the ability of a system to perform its task. Yet the framework commonly used for motif-detection averages over the local environments of all nodes. 
Therefore, it remains unclear whether motifs are overrepresented in the whole system or only in certain regions.

In this contribution, we overcome this limitation by mining \textit{node-specific} triad patterns. For every vertex, the abundance of each triad pattern is considered only in triads it participates in. We investigate systems of various fields and find that motifs are distributed highly heterogeneously. In particular we focus on the feed-forward loop motif which has been alleged to play a key role in biological networks.

\end{abstract}


%

\section{Introduction}

\blfootnote{Published in IEEE ICDMW 2014. \copyright 2014 IEEE. Personal use of this material is permitted. Permission from IEEE must be 
obtained for all other uses, in any current or future media, including 
reprinting/republishing this material for advertising or promotional purposes, creating new 
collective works, for resale or redistribution to servers or lists, or reuse of any copyrighted 
component of this work in other works.
}


The concept of networks has been successfully applied to model the interactions between entities in complex systems of various fields.
The topological structure of interactions among the constituents of complex many particle systems is intimately linked to system function and global system properties. A major branch of networks research aims to elucidate this link between structure and function.

Generally, pairwise relations between nodes, so-called dyads, are considered the fundamental building blocks of complex networks
. 
Erd\"{o}s-Renyi graphs~\cite{Erdos1959}, the configuration model~\cite{Molloy1995}, stochastic block models~\cite{Holland1983,Wang1987,Nowicki2001} and degree corrected block models~\cite{Karrer2010} all fall into the class of dyadic models. The basic assumption underlying dyadic models is that dyads are \emph{conditionally independent} given the model's parameters.

However, the assumption of dyadic independence as a general paradigm of network modeling seems questionable. For example, in a social context, the idea that the relation of Alice and Bob be independent from the relation of Alice and Charlie seems to go against experience, especially if the relation is of romantic type. Similarly, triadic closure, or the large clustering coefficient observed in many networks, hints at a dependence between the connections in a network. Generalizing these ideas, during the last decade the systematic study of higher order sub-network structure has attracted much attention~\cite{Milo2002,Milo2004,Sporns2004,Albert2004,berlingerio2009mining,rahman2014graft}. Subgraph patterns which occur significantly more often than expected at random, so called \textit{motifs}, have been hypothesized to serve as the acutal building blocks of many network topologies~\cite{Milo2002}.

The framework commonly applied to analyze local network structure detects an \textit{average} over- or underrepresentation of subgraph patterns for the whole system. Yet, especially for larger complex networks, there may be areas in which one structural pattern is of importance, whereas in different regions other patterns are relevant. The contribution of this work is to overcome this limitation by introducing a framework for \textit{node-specific} pattern mining. More specifically, instead of mining frequent subgraph patterns of the whole system, we investigate the neighborhood of every single node separately. I.e. for every node we consider only triads it participates in. In this paper, we further apply our new tool to multiple real-world data sets by analyzing their node-specific triad patterns. We show that for many systems their motifs are distributed highly heterogenously.

The remainder of this work will be organized as follows. In Section~\ref{sec:related} we briefly review work related to the analysis of networks in terms of their local substructure. In Section~\ref{sec:nspPatternMining} we describe our methodology of node-specific pattern mining. Results obtained from experiments on real datasets will be presented in Section~\ref{sec:results}.


\section{Related Work} \label{sec:related}

\begin{figure}
	\centering
	\includegraphics[width=0.48\textwidth]{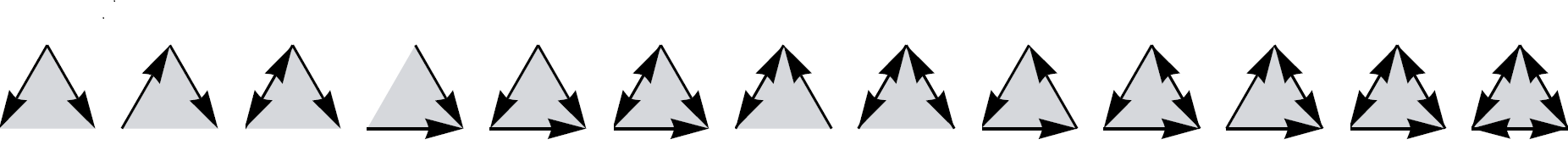}
	\caption{All 13 possible connected non-isomorphic triadic subgraphs (subgraph patterns) in directed unweighted networks.}
	\label{fig:triadPatterns}
\end{figure}

In the context of local subgraph analysis, most attention has been devoted to the investigation of triadic subgraphs \cite{Milo2002,Shen-Orr2002,Milo2004,Sporns2004,Albert2004}. Apart from node permutations, there are 13 distinct connected triad patterns in directed unweighted networks as shown in Figure~\ref{fig:triadPatterns}. It was found that certain patterns of third-order subgraphs occur significantly more frequently than expected in an ensemble of networks with the same degree distributions as the original network.

Over- and underrepresentation of pattern $i$ in a graph is quantified through a Z-score
\begin{equation}
	Z_{i} = \frac{N_{\text{original},i} - \left\langle N_{\text{rand},i}\right\rangle}{\sigma_{\text{rand},i}}.
	\label{eq:Zscore}
\end{equation}
$N_{\text{original},i}$ is the number of appearances of pattern $i$ over all possible 3-tuples of nodes in the original network. Sampling from the ensemble of randomized networks yields the average occurrence $\left\langle N_{\text{rand},i}\right\rangle$ of that pattern and the respective standard deviation $\sigma_{\text{rand},i}$. Thus, the Z-scores represent a measure for the significance of an over- or underrepresentation for each pattern $i$ shown in Figure~\ref{fig:triadPatterns}.

Hence, every network can be assigned a vector $\vec{Z}$ whose components comprise the Z-scores of all possible triad patterns. Significant patterns are referred to as 'motifs' \cite{Milo2002}. Further, one commonly refers to the normalized Z-vector as the 'significance profile', $\vec{SP} = \vec{Z} / \sqrt{ \sum_{i} Z_i^2}$. This normalization makes systems of different sizes comparable \cite{Milo2002}.

A multitude of real-world systems has been examined in terms of their triadic Z-score profiles \cite{Milo2004,Eom2006,Kaluza2010,Sakata2005} and it was found that systems with similar tasks tend to have similar Z-score profiles and thus exhibit the same motifs. Therefore, it was conjectured that their structural evolution may have been determined by the relevance of these motif patterns. Hence, motifs, rather than independent links, have been suspected to serve as the basic entities of complex network structure \cite{Milo2002}. In particular, the role of the 'feed-forward loop' pattern ($\motVIII$) has been discussed intensively in the field \cite{Shen-Orr2002,Mangan2003,Mangan2003_2,Alon2007}. It has been alleged to play a key role for systems to reliably perform information-processing tasks.

However, there has also been ongoing discussion about the expressive power of the subgraph-analysis described above \cite{Artzy-Randrup2004}. E.g. latent-variable models might offer an explanation for many of the observed motif distributions. The randomization employed in the subgraph-mining process ignores all mesoscopic structure, potentially present in the system. Hence, parts of the non-vanishing Z-scores may stem from such structure \cite{Reichardt2011,Beber}. 
Comparison of a network with block structure to a null model which does not account for such groups may result in over- or underrepresentations of patterns which are more than less artifacts of the mesoscopic structure. It was further shown that there are pairwise correlations in the Z-scores of the subgraph patterns in Figure~\ref{fig:triadPatterns}  which occur solely for statistical reasons and therefore have to be taken into account when interpreting the functional role of motif patterns \cite{Ginoza2010,Winkler2013}.

In the present work we aim to further unravel the role of triad motifs in complex networks. We introduce a methodology for node-specific pattern mining which allows us to localize the regions of a graph in which the instances of a motif predominantely appear. Thus, it is possible to identify and remove the nodes and links which eventually make a certain pattern a motif. This will enable future investigations to assess whether it is actually the presence of a motif which enables a system to perform its task or whether other structural aspects are more relevant.

\section{Node-Specific Triad Pattern Mining (NoSPaM)} \label{sec:nspPatternMining}

Based on the commonly applied subgraph-detection procedure we will now introduce the algorithm for \textbf{No}de-\textbf{S}pecific \textbf{Pa}ttern \textbf{M}ining  (\textsc{NoSPaM}). For every node $\alpha$ in a graph \textsc{NoSPaM} evaluates the abundance of all structural patterns in $\alpha$'s neighborhood, i.e. patterns $\alpha$ participates in, and compares it to the expected frequency of occurrence in a randomized ensemble of the network under investigation. In the latter, both individual in- and out degrees of all nodes, and the number of unidirectional and bidirectional links are the same as in the original network.

The \textsc{NoSPaM} framework can be realized for patterns comprised of an arbitrary number of nodes $n$. The detailed algorithmic implementation of \textsc{NoSPaM}$_n$ is then dependent on the pattern size. For the rest of this work we will focus on triad patterns ($n=3$). However, a generalization to higher orders is straightforward. Although, of course, with increasing $n$, the number of non-isomorphic subgraphs increases rapidly.

\begin{figure}
	\centering
	\includegraphics[width=0.490\textwidth]{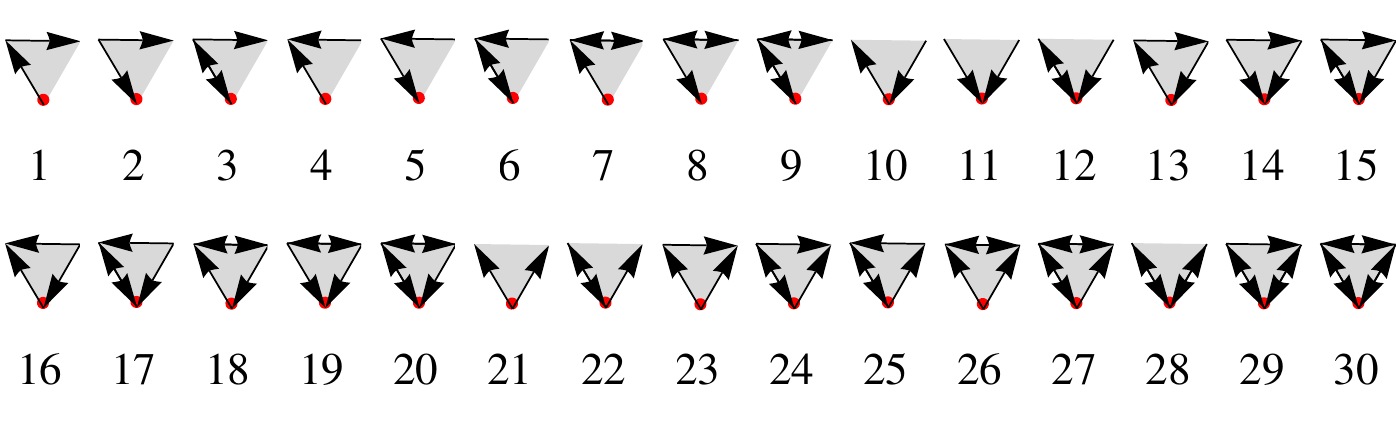}
	\caption{All possible connected, nonisomorphic triadic subgraph patterns in terms of a distinct node (here: lower node).}
	\label{fig:dir_triadLocalPatterns_connected}
\end{figure}

We aim to evaluate the abundance of triad patterns from a node $\alpha$'s point of view. Therefore, the symmetry of most patterns of Figure~\ref{fig:triadPatterns} is now broken and the number of connected \textit{node-specific} triad patterns increases to 30 rather than 13. They are shown in Figure~\ref{fig:dir_triadLocalPatterns_connected}.
To understand the increase in the number of patterns consider the oridinary subgraph~$\motV$. From the perspective of one particular node, this pattern splits into the three node-specific triad patterns 1, 5, and~10 in Figure~\ref{fig:dir_triadLocalPatterns_connected}.

Of course, some patterns are included in others, e.g. pattern~1 in pattern~3. It shall be emphasized that, in order to avoid biased results, we do not double count, i.e. an observation of pattern~3 will only increase its corresponding count and not the one associated with pattern~1.

For every node $\alpha$ in a graph, \textsc{NoSPaM}$_3$ will now compute Z-scores for each of the 30 node-specific patterns $i$ shown above:
\begin{equation}
	Z_{i}^{\alpha} = \frac{N_{\text{original},i}^{\alpha} - \left\langle N_{\text{rand},i}^{\alpha}\right\rangle}{\sigma_{\text{rand},i}^{\alpha}}.
	\label{eq:nodeZscore}
\end{equation}
$N_{\text{original},i}^{\alpha}$ now is the number of appearances of pattern $i$ in the triads where node $\alpha$ participates in. Accordingly, $\left\langle N_{\text{rand},i}^{\alpha}\right\rangle$ is the expected frequency of pattern $i$ in the triads where node $\alpha$ is part of in the randomized ensemble. $\sigma_{\text{rand},i}^{\alpha}$ is the corresponding standard deviation.

\begin{figure}
	\centering
	\subfigure[][]{
		\centering
		\includegraphics[width=0.270\textwidth]{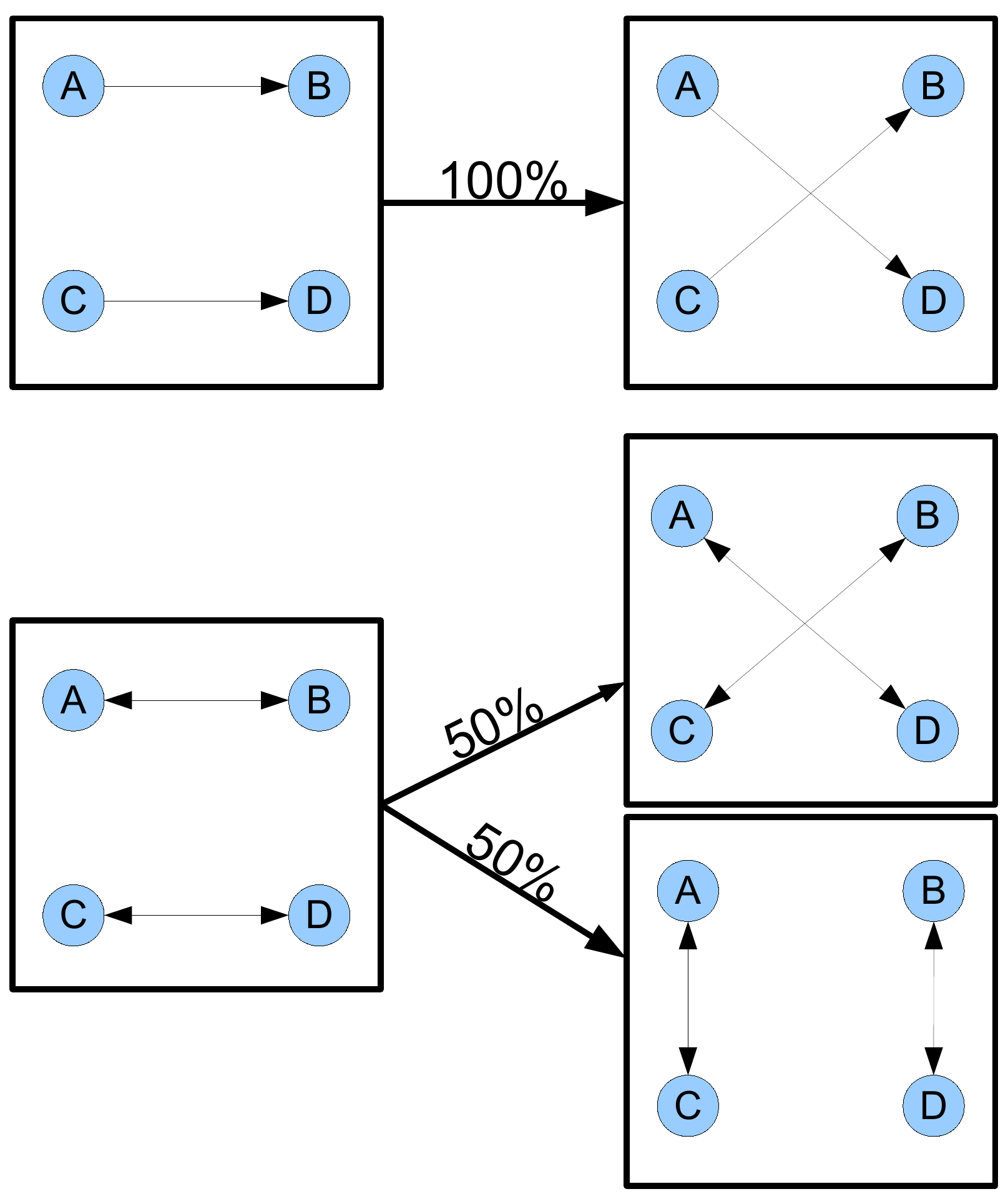}
	\label{fig:randomization1}
	}
	\hspace{0.3cm}
	\subfigure[][]{
		\includegraphics[width=0.270\textwidth]{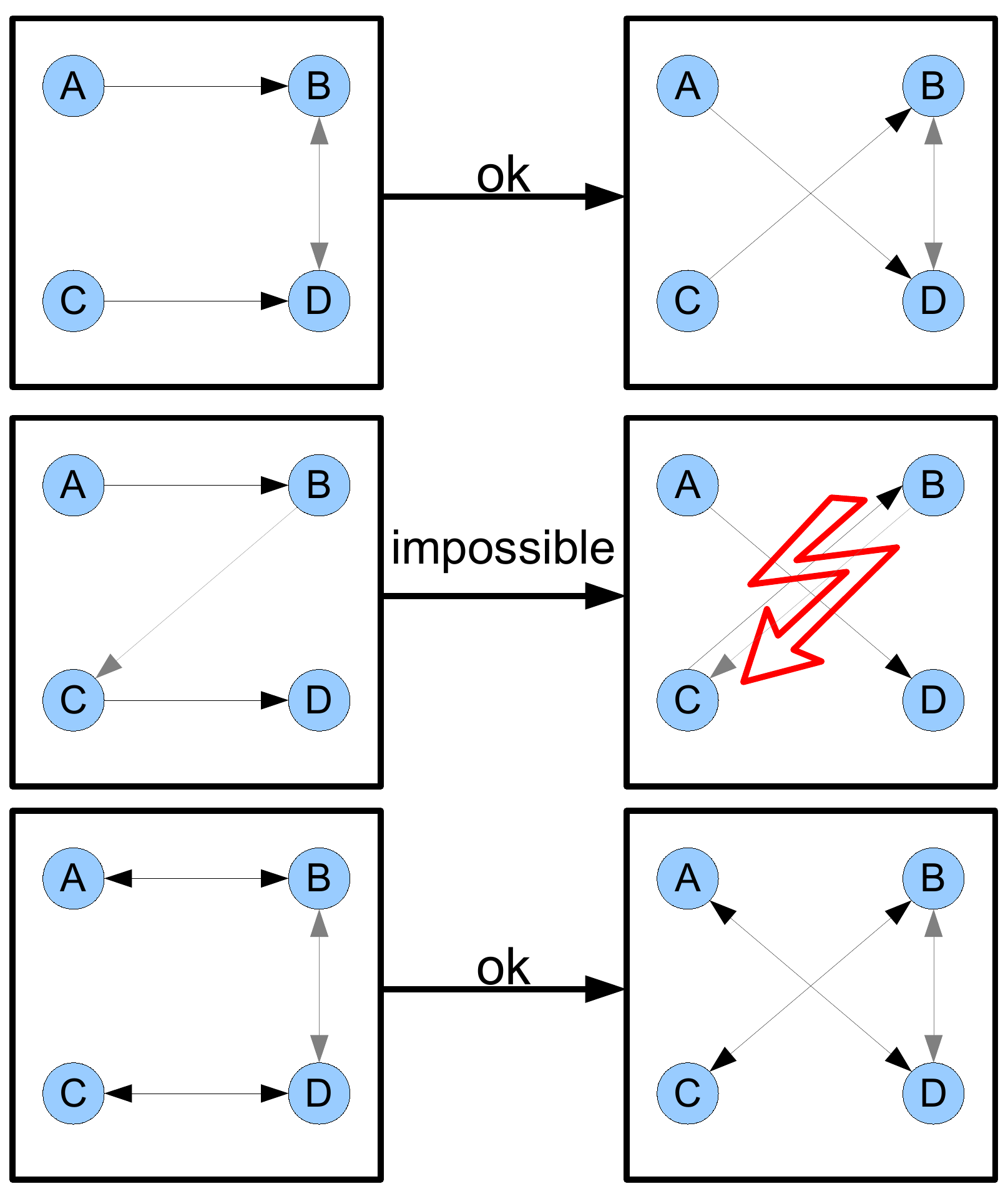}
	\label{fig:randomization2}
	}
	\caption{Microscopic link-switchings performed to generate the randomized ensembles. (a)~illustrates the switching rules. (b)~shows cases where additional links affect the randomization process: only if these links interfer with the rewiring of a particular edge (middle) the step cannot be performed.}
	\label{fig:randomization}
\end{figure}

The algorithm for the node-specific triad pattern mining will now be presented in three parts. Algorithm~\ref{alg:randomization} describes the randomization process which is the same as the shuffling done for ordinary subgraph mining. The microscopic steps performed to randomize a network are illustrated in Figure~\ref{fig:randomization}.

\begin{algorithm}
  \caption{Degree-preserving randomization of a graph
    \label{alg:randomization}}
  \begin{algorithmic}
	\Function{Randomize}{Graph $\mathcal{G}$, no. of required steps}
    \State success = 0
    \While{success $<$ number of required rewiring steps}
    \State pick a random link $e_1 \in \mathcal{G}$
    \If{$e_1$ is unidirectional}
    	\State pick a 2nd unidirectional link $e_2 \in \mathcal{G}$ at random
    \Else
    	\State pick a 2nd bidirectional link $e_2 \in \mathcal{G}$ at random
    \EndIf
    \If{$e_1$ and $e_2$ do not share a node}
    \State rewire according to the rules in Figure~\ref{fig:randomization1}
    \State success++
    	\If{one of the new links already exists}
    		\State undo the rewiring
    		\State success$--$
    	\EndIf
    \EndIf
    
    \EndWhile
    \State \Return randomized instance of $\mathcal{G}$
   \EndFunction
  \end{algorithmic}
\end{algorithm}

\noindent The number of rewiring steps should be chosen proportionally to the number of links in the graph. Since all randomization steps preserve individual node degrees and the number of unidirectional and bidirectional links, these quantities are also conserved on the macroscopic level. It shall be mentioned that alternative randomization methods exist: exponential random graph models (ERGMs) allow for a faster generation of randomized networks~\cite{DeBie2010}, however they come with the limitation to fix only the expectation for individual node degrees, not necessarily the actual values. For applications of the methodology to big data ERGMs may serve as an alternative to generate the random null models, yet going along with a loss of accuracy.

\begin{algorithm}
  \caption{Counting of node-specific triad patterns
    \label{alg:nspTriadCounting}}
  \begin{algorithmic}
	\Function{NspPatternCounter}{Graph $\mathcal{G}$}
		\State $\mathcal{N}$: $N \times 30$-dimensional array storing the pattern counts
		\State \qquad for every node of $\mathcal{G}$
    \For{every edge $e \in \mathcal{G}$}
    	\State $i, j$ $\leftarrow$ IDs of $e$'s nodes with $i < j$
    	\State $\mathcal{C} \leftarrow \left\{\right\}$ be list of candidate nodes to form triad
    	\State \qquad \qquad patterns including $e$
    	\State $\mathcal{C} \leftarrow $ all neighbors of $i$
    	\State $\mathcal{C} \leftarrow $ all neighbors of $j$
    	\For{all $c \in \mathcal{C}$}
    		\If{$i+j < $ sum of IDs of all other \textit{connected}\\ \qquad \qquad \qquad \qquad \qquad dyads in triad $(ijc)$}
    			\State increase the counts in $\mathcal{N}$ for $i$, $j$, and $c$ for 
    			\State their respective node-specific patterns
    		\EndIf
    	\EndFor
    \EndFor
    \State \Return $\mathcal{N}$
   \EndFunction
  \end{algorithmic}
\end{algorithm}

Algorithm~\ref{alg:nspTriadCounting} performs the counting process for the appearances of triad patterns in a graph. Because it is computationally expensive to test all triads in the system ($\mathcal{O}\left(N^3\right)$), we rather iterate over pairs of adjacent edges in the graph. Since real-world networks are usually sparse, this is much more efficient.


\begin{algorithm}
  \caption{Node-specific triad pattern mining (\textsc{NoSPaM}$_3$)
    \label{alg:nospam3}}
  \begin{algorithmic}
	\Function{NoSPaM}{Graph $\mathcal{G}$, \# required rewiring steps, \#~randomized instances}
		\State $\mathcal{N}_\text{original} \leftarrow$ \textsc{NspPatternCounter}($\mathcal{G}$)
		\State $\mathcal{N}_\text{rand} \leftarrow \left\{\right\}$
		\State $\mathcal{N}_\text{sq,rand} \leftarrow \left\{\right\}$
		\For{\# randomized instances}
			\State $\mathcal{G} \leftarrow $ \textsc{Randomize}($\mathcal{G}$, \# required rewiring steps)
			\State counts $\leftarrow$ \textsc{NspPatternCounter}($\mathcal{G}$) 
			\State $\mathcal{N}_\text{rand} \leftarrow \mathcal{N}_\text{rand} +$ counts
			\State $\mathcal{N}_\text{sq,rand} \leftarrow \mathcal{N}_\text{sq,rand} +$ counts $*$ counts
		\EndFor
		\State $\mathcal{N}_\text{rand} \leftarrow \mathcal{N}_\text{rand} / (\# \text{randomized instances})$ 
		\State $\mathcal{N}_\text{sq,rand} \leftarrow \mathcal{N}_\text{sq,rand} / (\# \text{randomized instances})$ 
		\State $\mathcal{\sigma}_\text{rand} \leftarrow \sqrt{\mathcal{N}_\text{sq,rand}-(\mathcal{N}_\text{rand}*\mathcal{N}_\text{rand})}$ 
		\State $\mathcal{Z} \leftarrow (\mathcal{N}_\text{original} - \mathcal{N}_\text{rand})/\mathcal{\sigma}_\text{rand}$ 
		\State \Return $\mathcal{Z}$
   \EndFunction
  \end{algorithmic}
\end{algorithm}

Using the functions defined in Algorithm~\ref{alg:randomization} and Algorithm~\ref{alg:nspTriadCounting}, we can eventually formulate the routine of our new methodology of node-specific triad pattern mining (\textsc{NoSPaM}$_3$). Algorithm~\ref{alg:nospam3} describes its formalism. It computes the node-specific Z-scores as defined in Equation~\ref{eq:nodeZscore}. All operations on arrays in Algorithm~\ref{alg:nospam3} are performed elementwise.

\textbf{Performance:} The computational cost of Algorithm~\ref{alg:randomization}, $C_1$, scales with the number of required randomization steps per instance, which should be chosen proportionally to the number of edges $E$ in the graph $\mathcal{G}$, i.e. $C_1 = \mathcal{O}\left(E\right)$.

Algorithm~\ref{alg:nspTriadCounting} iterates over all edges of $\mathcal{G}$ and their adjacent edges. Therefore, it is $C_2 = \mathcal{O}\left(E \cdot k_\text{max}\right) \leq \mathcal{O}\left(E^2\right)$ where $k_\text{max}$ is the maximum node degree in $\mathcal{G}$. In real-world networks $k_\text{max}$ is usually much smaller than $E$.

Finally, the total computational cost of \textsc{NoSPaM}$_3$ (Algorithm~\ref{alg:nospam3}) depends on the desired number of randomized network instances $I$. Algorithm~\ref{alg:nspTriadCounting} is invoked $(1+I)$~times, Algorithm~\ref{alg:randomization} is invoked $I$~times. Hence, the total computational cost is
\begin{equation}
	C_{\textsc{NoSPaM}_3} = \mathcal{O}\left(I \cdot E \cdot k_\text{max} \right).
\label{eq:compCost}
\end{equation}
Further, it shall be mentioned that $\textsc{NoSPaM}_3$ is parallelizeable straightforwardly since the evaluations in terms of the randomized network instances can be executed independently of each other.


\section{Results} \label{sec:results}

We will now present results obtained from the application of \textsc{NoSPaM}$_3$ to peer-reviewed real-world datasets. All networks are directed and edges are treated as unweighted. An implementation of the pattern-mining program is made publically available online~\cite{WinklerWebsite}.

\textbf{Yeast transcriptional}~\cite{AlonWebpage,Costanzo2001}\textbf{:} 688 nodes, 1,079 edges. Transcriptional network of the yeast S.~Cerevisiae. Nodes are genes, edges point from regulating genes to regulated genes. It is not distinguished between activation and repression.

\textbf{E. Coli transcriptional}~\cite{AlonWebpage,Mangan2003}\textbf{:} 423 nodes, 519 edges. Nodes are operons, each edge is directed from an operon that encodes a transcription factor to an operon that it directly regulates (an operon is one or more genes transcribed on the same mRNA).

\textbf{Neural network of C. Elegans}~\cite{Varshney2011,wormatlas}\textbf{:} 279 nodes, 2,194 edges. Nodes are the neurons of the largest connected component in the somatic nervous system. Edges describe the chemical synapses between the neurons.

\textbf{Scientific citations}~\cite{Gehrke:2003:OKC:980972.980992,Leskovec:2005:GOT:1081870.1081893}\textbf{:} 27,700 nodes, 352,807 edges. Nodes are high-energy physics papers on the arXiv, submitted between January 1993 and April 2003. Edges from node A to B indicate that paper A cites paper B. Although it may seem unintuitive, there are papers citing each other.

\textbf{Political blogs}~\cite{NewmanData,Adamic2005}\textbf{:} 1,224 nodes, 19,025 edges. Largest connected component of a network where the nodes are political blogs. Edges represent links between the blogs recorded over a period of two months preceding the 2004 US Presidential election.

\textbf{French book}~\cite{AlonWebpage,Milo2004}\textbf{:} 8,325 nodes, 24,295 edges. Word-adjacency network of a French book. Nodes are words, an edge from node A to B indicates that word B directly follows word A at least once in the text.

\textbf{Spanish book}~\cite{AlonWebpage,Milo2004}\textbf{:} 11,586 nodes, 45,129 edges. Word-adjacency network of a Spanish book.

\subsection{Node-specific vs. ordinary triadic Z-score profiles}

\begin{figure*}
\newcommand{\w}{0.47}
	\centering
	\subfigure[][Yeast transcriptional]{
		\includegraphics[width=\w\textwidth]{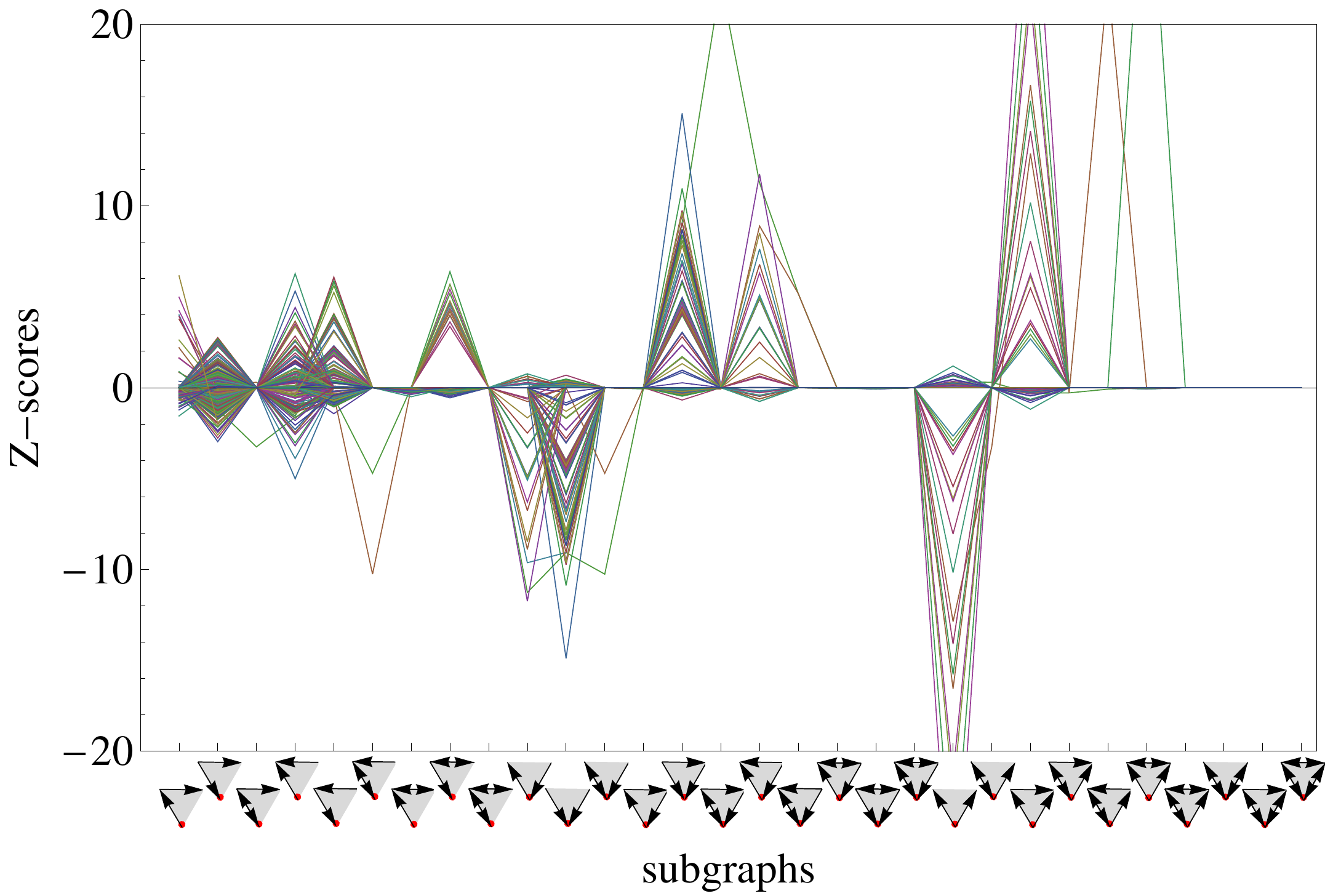}
		\label{fig:yeastInter_st_nspZ_normalized_all}
	}
	\subfigure[][E. Coli transcriptional]{
		\includegraphics[width=\w\textwidth]{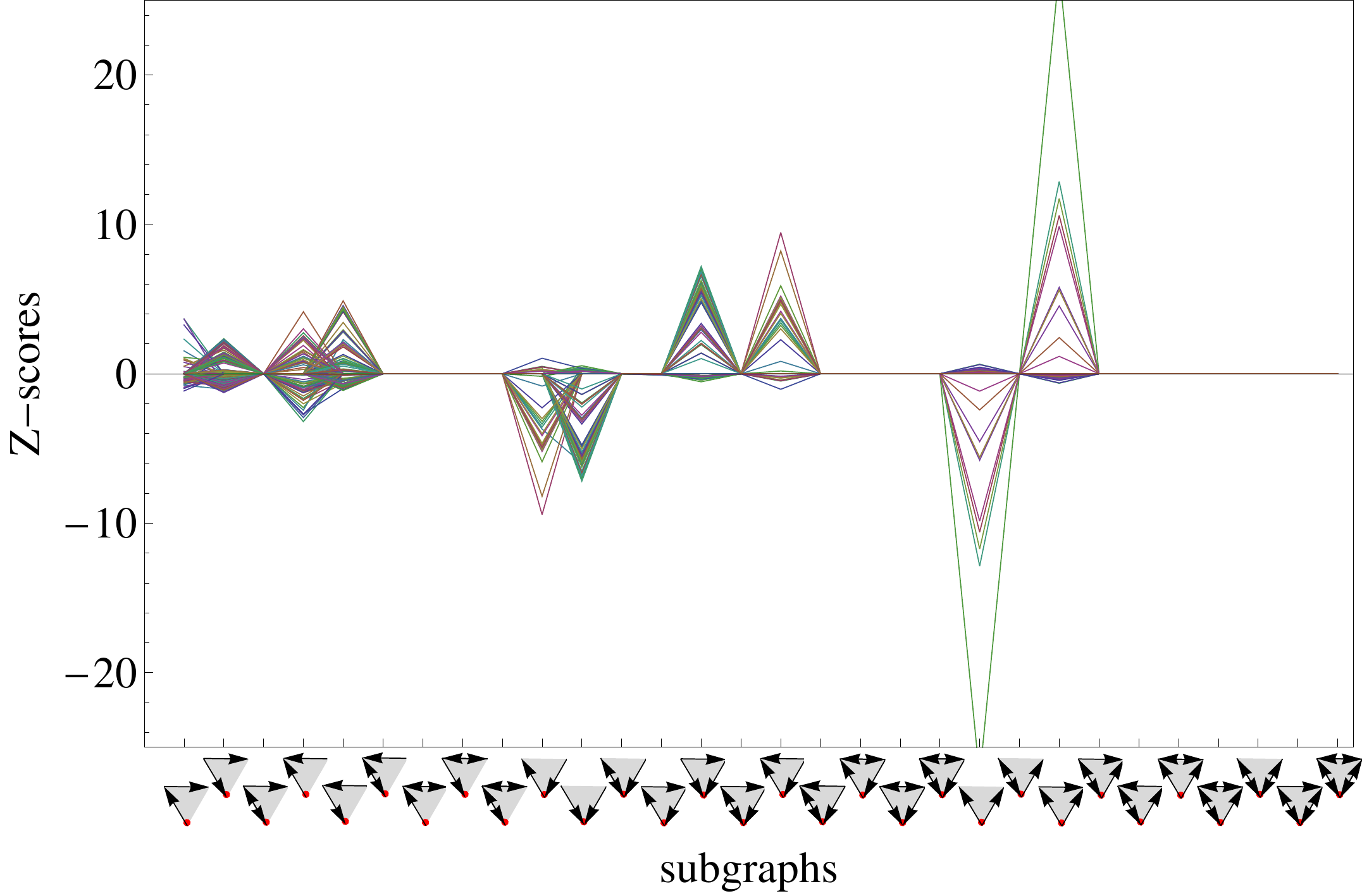}
		\label{fig:coli1_1Inter_nspZ_all}
	}\\
	\subfigure[][Scientific citations]{
		\includegraphics[width=\w\textwidth]{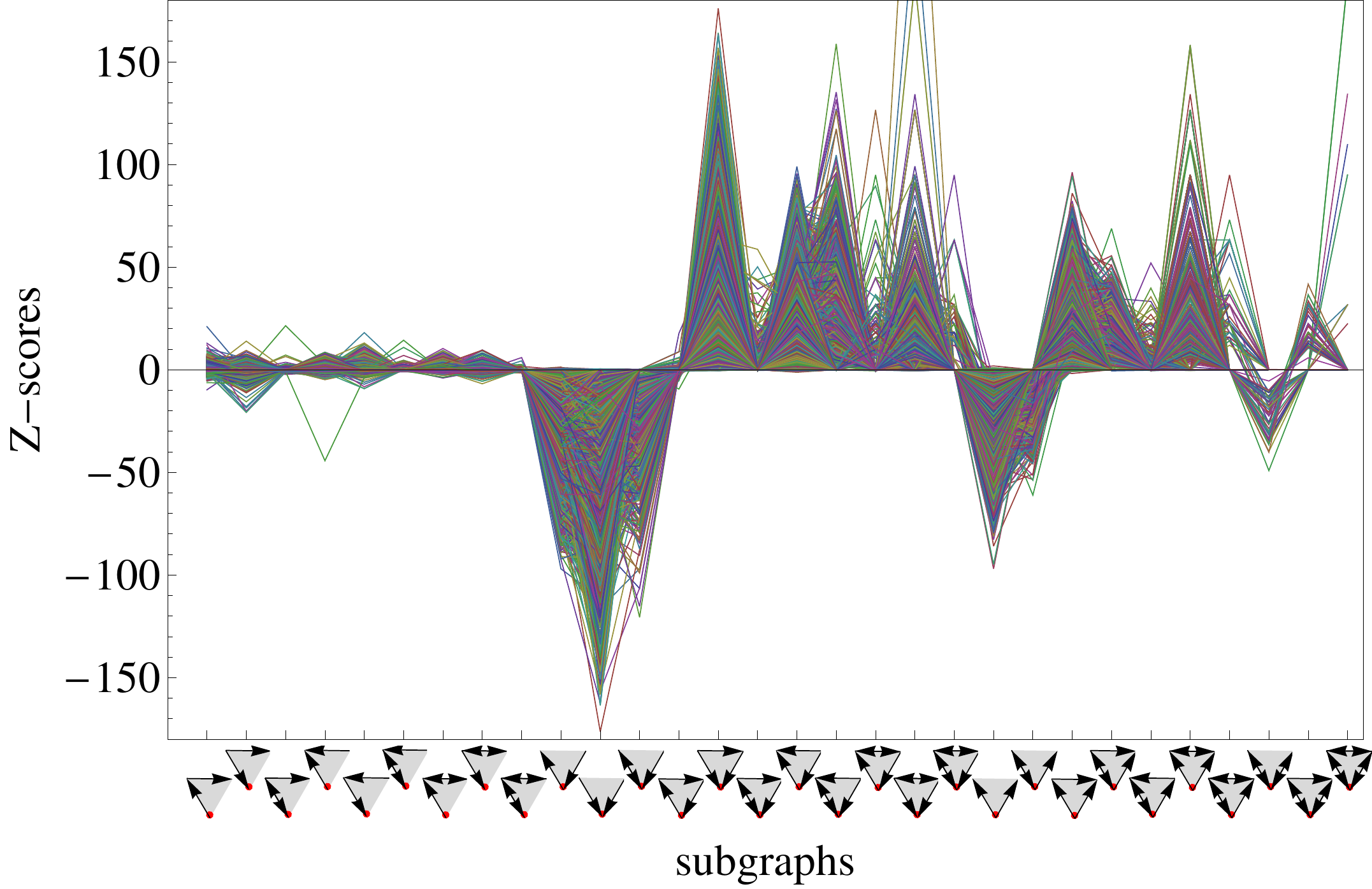}
		\label{fig:Cit-HepTh_nspZ_all}
	}
	\subfigure[][Political blogs]{
		\includegraphics[width=\w\textwidth]{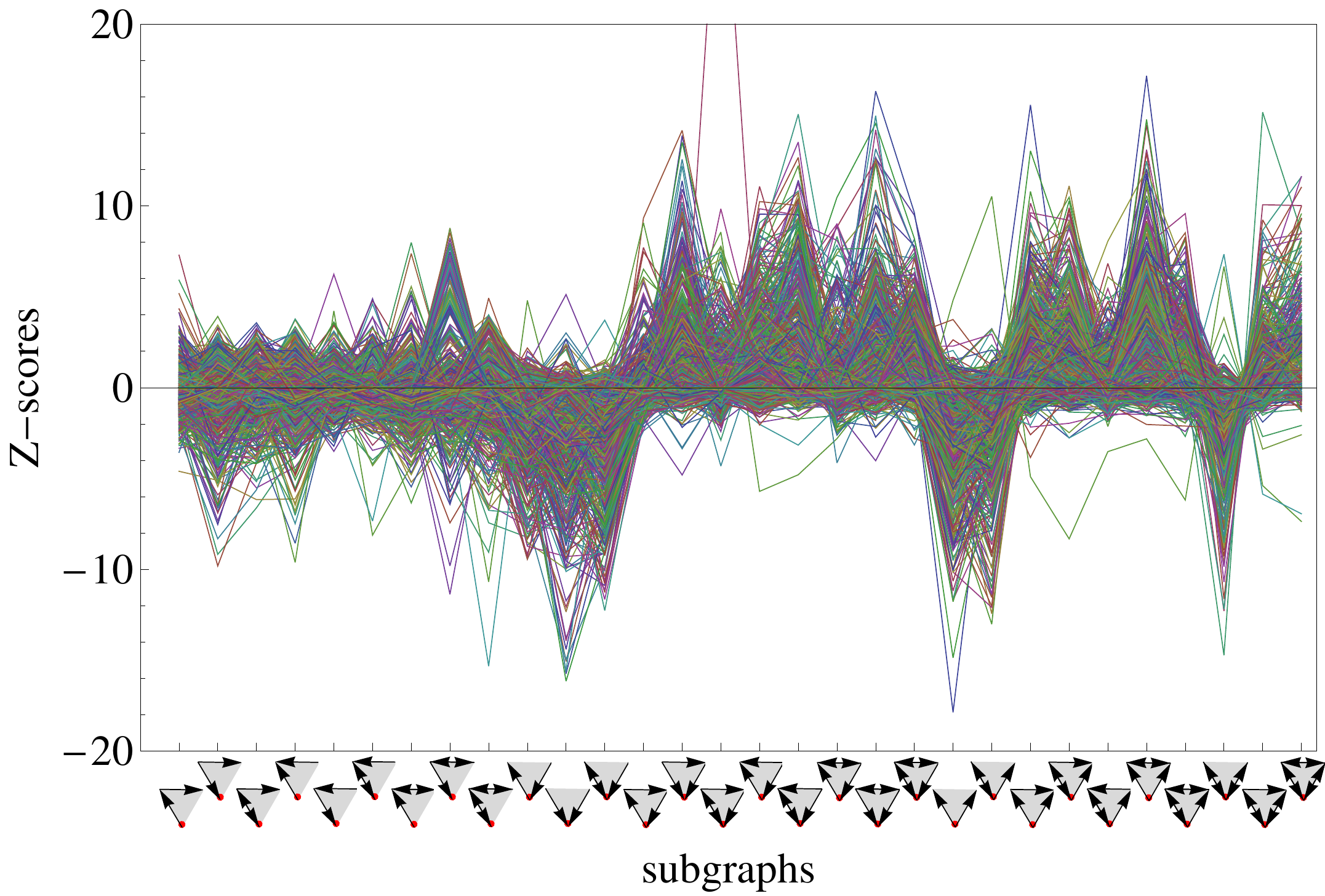}
		\label{fig:allZ_polblogs_Inter}
	}\\
	\subfigure[][French book]{
		\includegraphics[width=\w\textwidth]{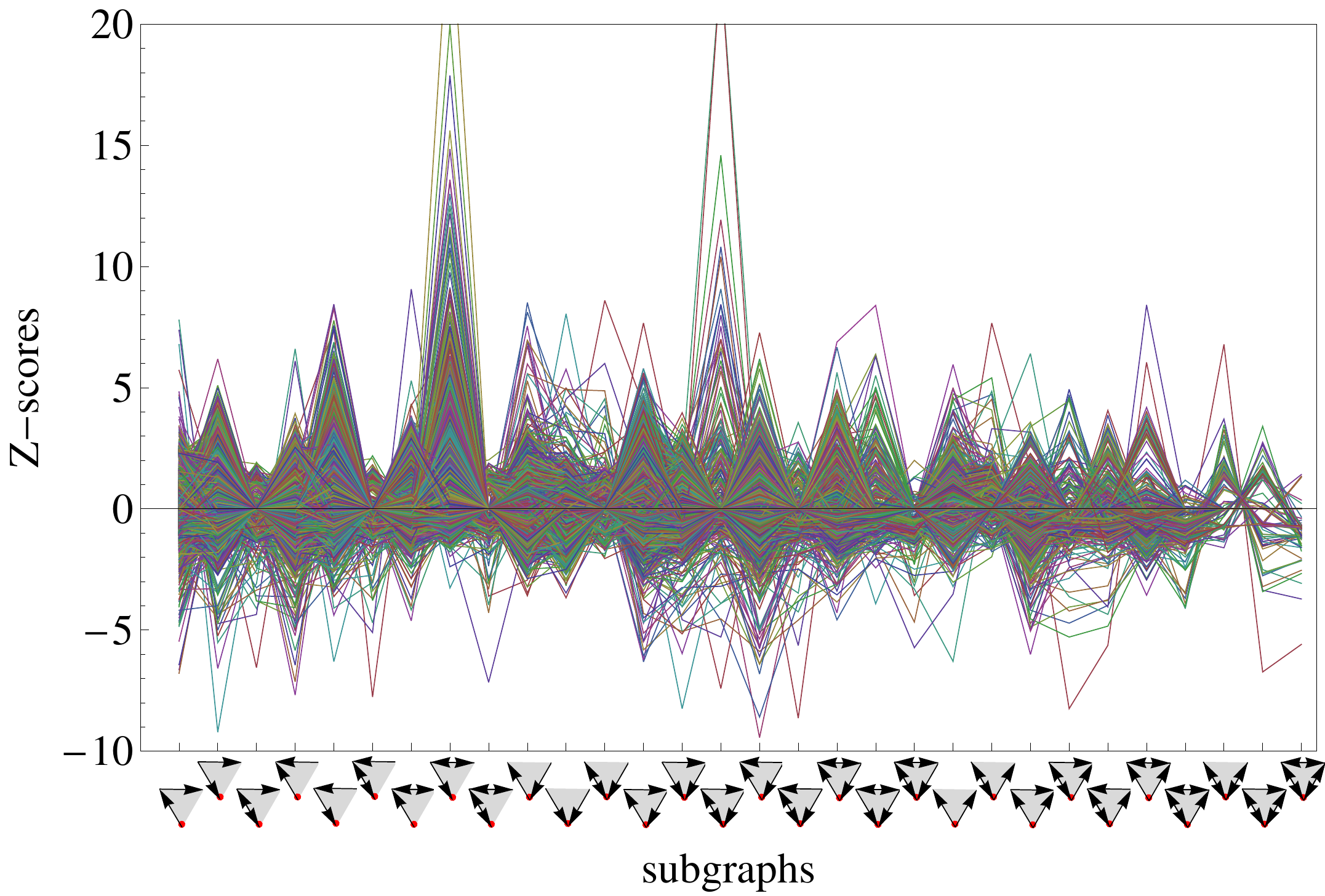}
		\label{fig:allZ_frenchBookInter}
	}
	\subfigure[][Spanish book]{
		\includegraphics[width=\w\textwidth]{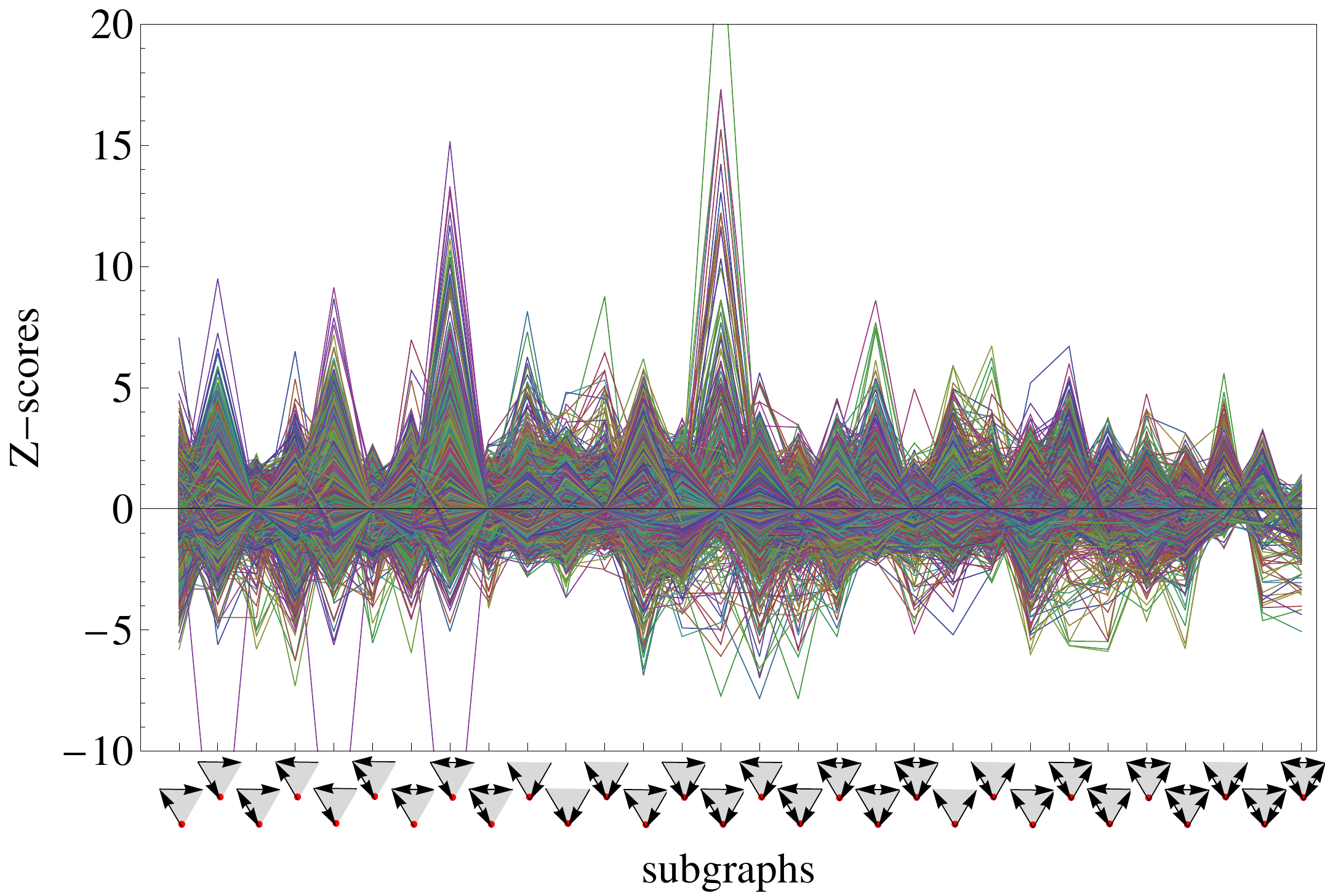}
		\label{fig:allZ_spanishBookInter}
	}
	\caption{Node-specific Z-score profiles of various real-world networks: transcriptional networks of a)~the yeast S. Cerevisiae~\cite{AlonWebpage,Costanzo2001} and b)~E. Coli~\cite{AlonWebpage,Mangan2003}, socially related networks such as c)~the citations of scientific papers~\cite{Gehrke:2003:OKC:980972.980992,Leskovec:2005:GOT:1081870.1081893} and d)~hyperlinks between political blogs~\cite{NewmanData,Adamic2005}, and word-adjacency networks of e)~French books~\cite{AlonWebpage,Milo2004} and f)~Spanish books~\cite{AlonWebpage,Milo2004}. The node-specific patterns on the horizontal axis are oriented the way that the node under consideration is the lower one.}
	\label{fig:nspZ_profiles_real1}
\end{figure*}

Figure~\ref{fig:nspZ_profiles_real1} shows the node-specific triadic Z-score profiles for various systems. We used at least five switches per edge (on average) for every iteration and averaged over 1000 iterations. Note that there is one curve for every node in the graph. The node-specific patterns on the horizontal axis are oriented the way that the node under consideration is the lower one.

We find that systems from similar fields have similar node-specific triadic Z-score profiles. Figures~\ref{fig:yeastInter_st_nspZ_normalized_all} and~\ref{fig:coli1_1Inter_nspZ_all} show transcriptional networks, Figures~\ref{fig:Cit-HepTh_nspZ_all} and~\ref{fig:allZ_polblogs_Inter} show data from a social context, specifically the citation network of high-energy physicists and the network of hyperlinks between political blogs. Figures~\ref{fig:allZ_frenchBookInter} and~\ref{fig:allZ_spanishBookInter} show word-adjacency networks in French and Spanish language, respectively. The observation that systems from a similar context exhibit similar local structural aspects indicate that the latter are strongly linked to the systems' function.

The fact that \textsc{NoSPaM}$_3$ provides localized data enables us to identify the areas of a graph where certain subgraph patterns primarily occur. Particularly, it allows us to test whether motifs of a system are overabundant throughout the entire network or if they are restricted to limited regions or the proximity of few nodes. In order to explore this issue, for each node, we will map its node-specific Z-scores to a score for the regular triad patterns (shown in Figure~\ref{fig:triadPatterns}). This will be realized by taking the mean over the Z-scores of all node-specific triad patterns corresponding to a regular triad pattern. The mapping is shown in Table~\ref{tab:patternMapping}. Hence, we obtain a 13-dimensional mapped node-specific Z-score profile for every node in a graph.

\begin{table*}
	\centering
	\caption{Mapping of node-specific triad patterns to their regular triad patterns.}
		\begin{tabular}{c || c | c | c | c | c| c | c | c | c | c | c | c | c}
				 Node-specific 	&    								& \large$\nmotX$  & \large$\nmotXXII$& 									& \large$\nmotXXIII$& 									& \large$\nmotXII$& 								& 									& \large$\nmotXXV$	& 									& \large$\nmotXXIX$\\
				 triad patterns & \large$\nmotXXI$  & \large$\nmotV$  & \large$\nmotVIII$& \large$\nmotXI$	& \large$\nmotXVI$	& \large$\nmotXIV$	&	\large$\nmotVII$& \large$\nmotXVIII$& 								& \large$\nmotXVIII$& \large$\nmotXXVI$	& \large$\nmotXXVII$\\
				 								& \large$\nmotII$ 	& \large$\nmotI$  & \large$\nmotIII$& \large$\nmotIV$		& \large$\nmotXIV$	& \large$\nmotXIX$	& \large$\nmotVI$	& \large$\nmotIX$	& \large$\nmotXIII$	& \large$\nmotXV$		& \large$\nmotXVII$	& \large$\nmotXX$	& \large$\nmotXXX$\\
				\hline
				\hline
				Regular&   \,\,\,\,\,\, & \,\,\,\,\,\, &  \,\,\,\,\,\, & \,\,\,\,\,\, &\,\,\,\,\,\, &  \,\,\,\,\,\, & \,\,\,\,\,\, & \,\,\,\,\,\, &  \,\,\,\,\,\, &  \,\,\,\,\,\, &  \,\,\,\,\,\, &  \,\,\,\,\,\, & \,\, \\
				triad patterns & \large$\motIV$   	&\large$\motV$   	& \large$\motVI$	& \large$\motVII$   &\large$\motVIII$   & \large$\motIX$   	& \large$\motX$		& \large$\motXI$	& \large$\motXII$		& \large$\motXIII$	& \large$\motXIV$	& \large$\motXV$	& \large$\motXVI$
		\end{tabular}
	\label{tab:patternMapping}
\end{table*}

\begin{figure*}
\newcommand{\w}{0.35}
	\centering
	\subfigure[][Yeast transcriptional]{
		\includegraphics[width=\w\textwidth]{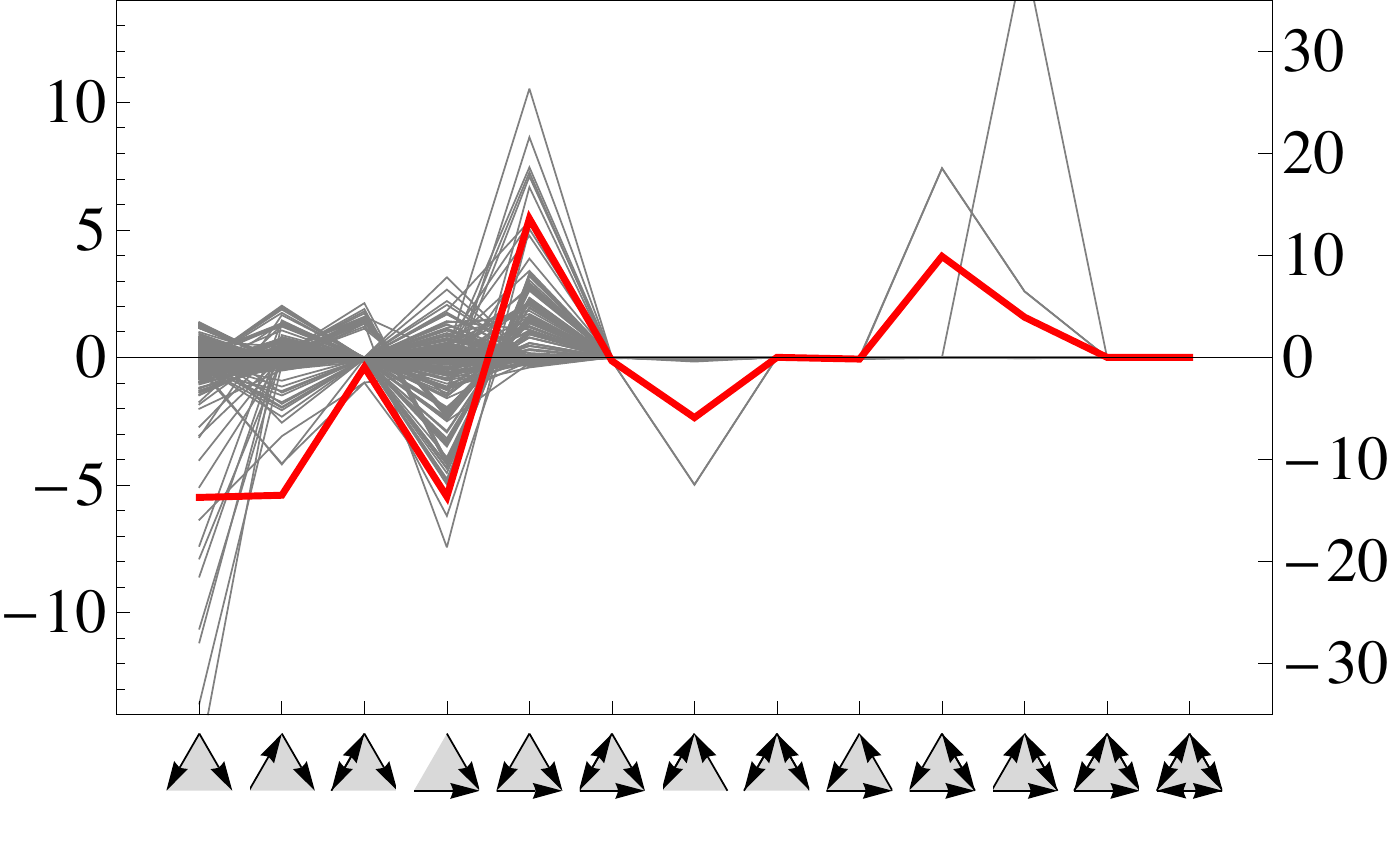}
		\label{fig:nspZ_profiles_mapped_mean_yeastInter}
	}
	\subfigure[][E. Coli transcriptional]{
		\includegraphics[width=\w\textwidth]{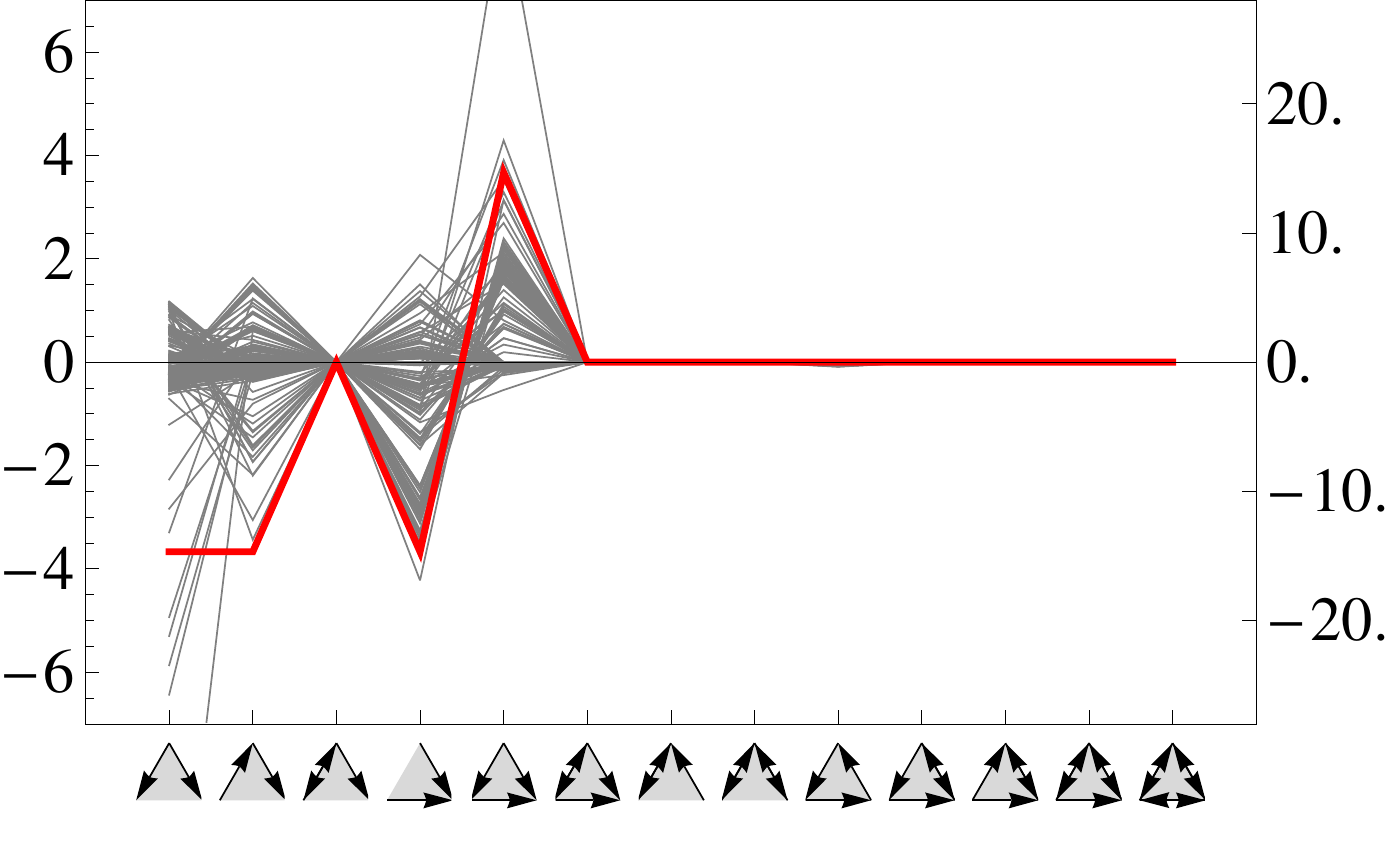}
		\label{fig:nspZ_profiles_mapped_mean_coli1_1Inter}
	}\\
	\subfigure[][C.Elegans neural network]{
		\includegraphics[width=\w\textwidth]{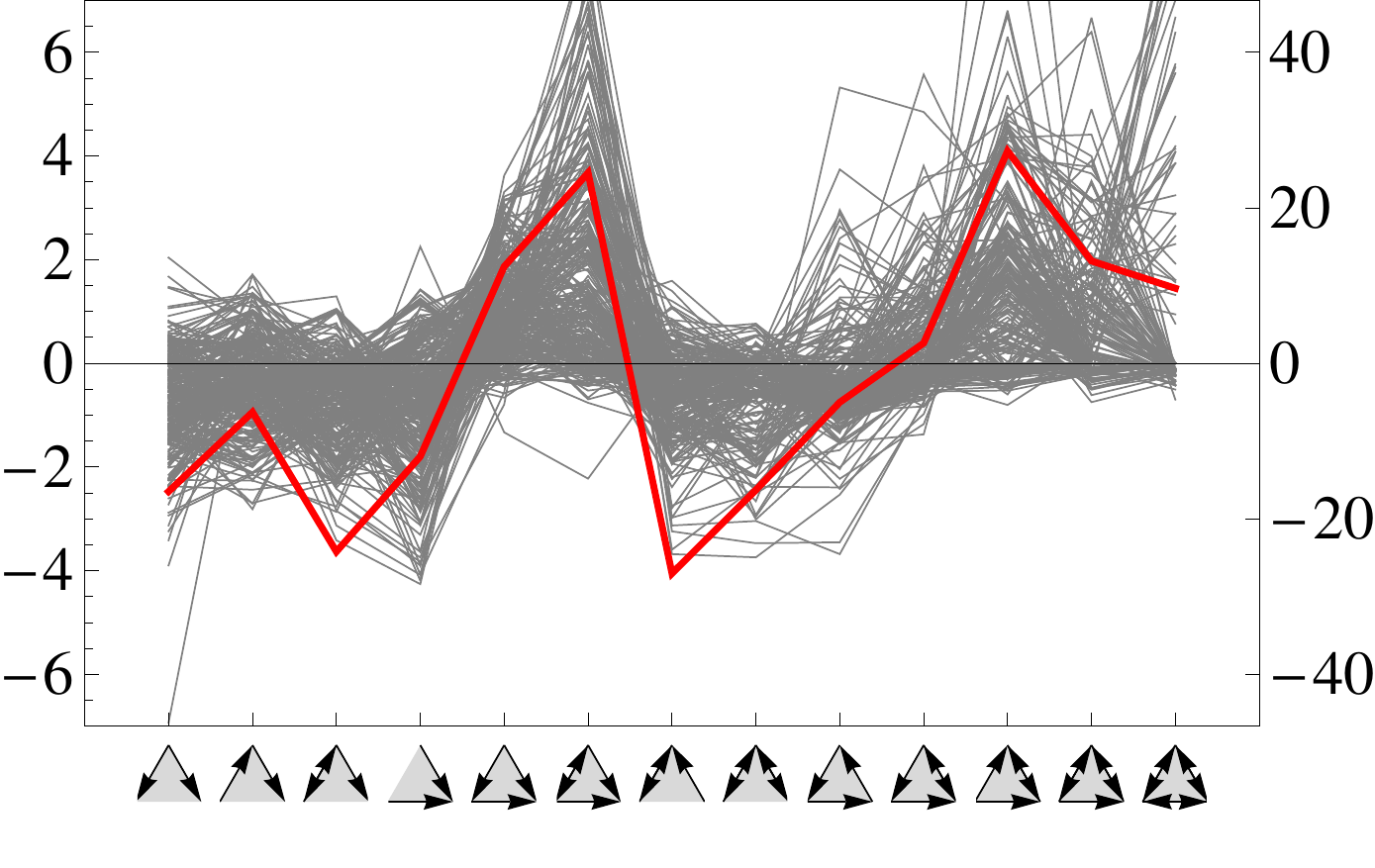}
		\label{fig:nspZ_profiles_mapped_mean_celegansneural}
	}
	\subfigure[][Political blogs]{
		\includegraphics[width=\w\textwidth]{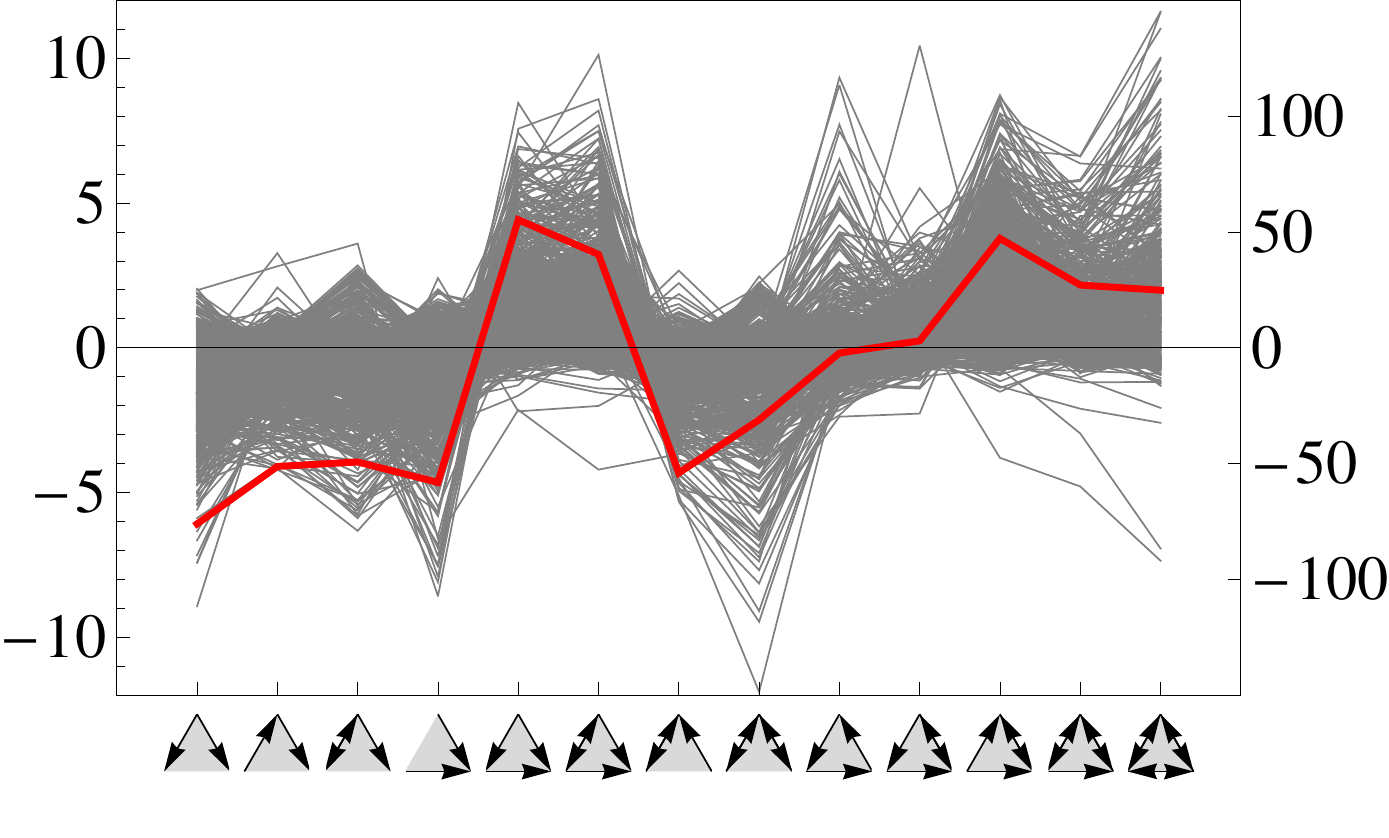}
		\label{fig:nspZ_profiles_mapped_mean_polblogs_Inter}
	}\\
	\subfigure[][French book]{
		\includegraphics[width=\w\textwidth]{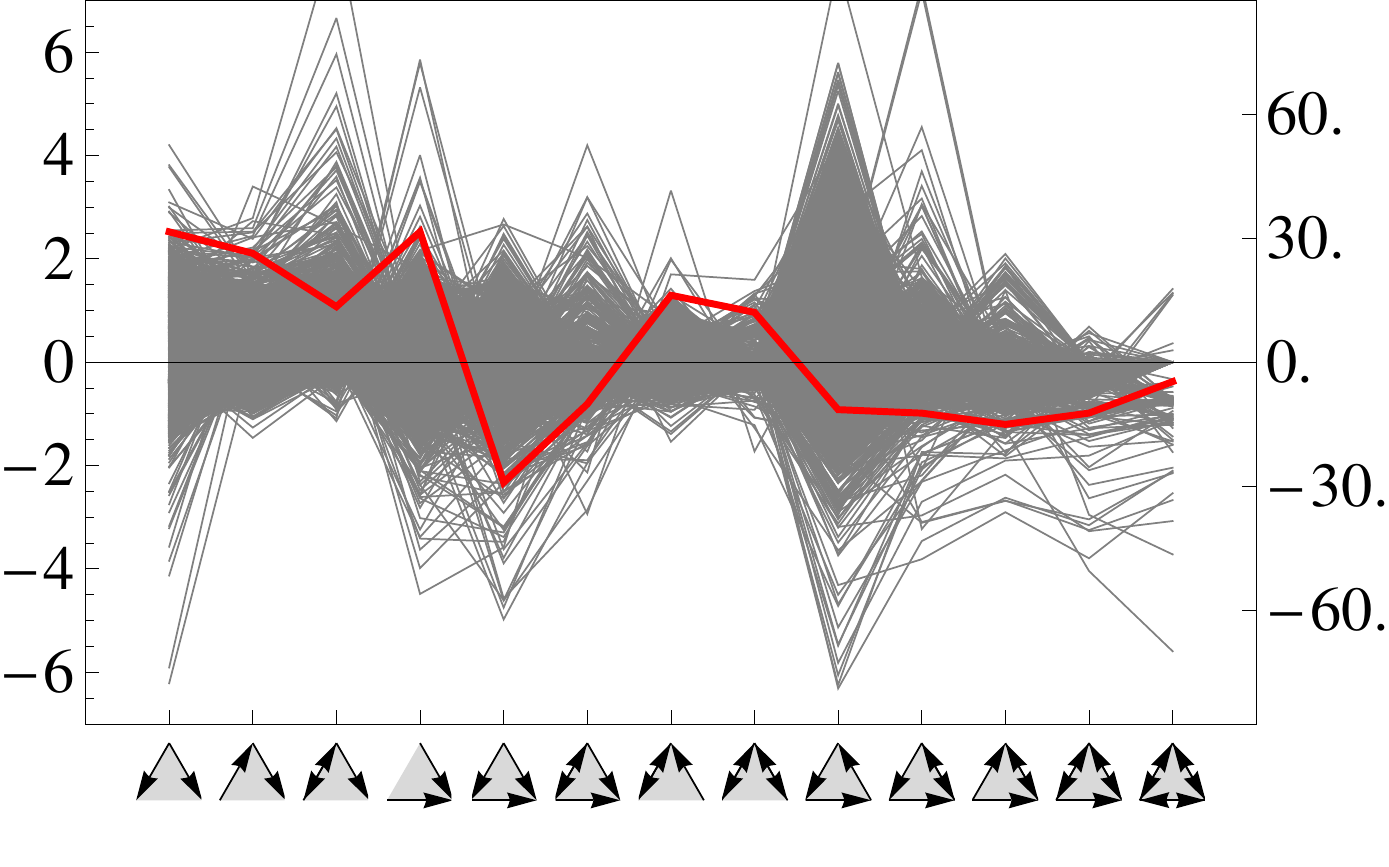}
		\label{fig:nspZ_profiles_mapped_mean_frenchBookInter}
	}
	\subfigure[][Spanish book]{
		\includegraphics[width=\w\textwidth]{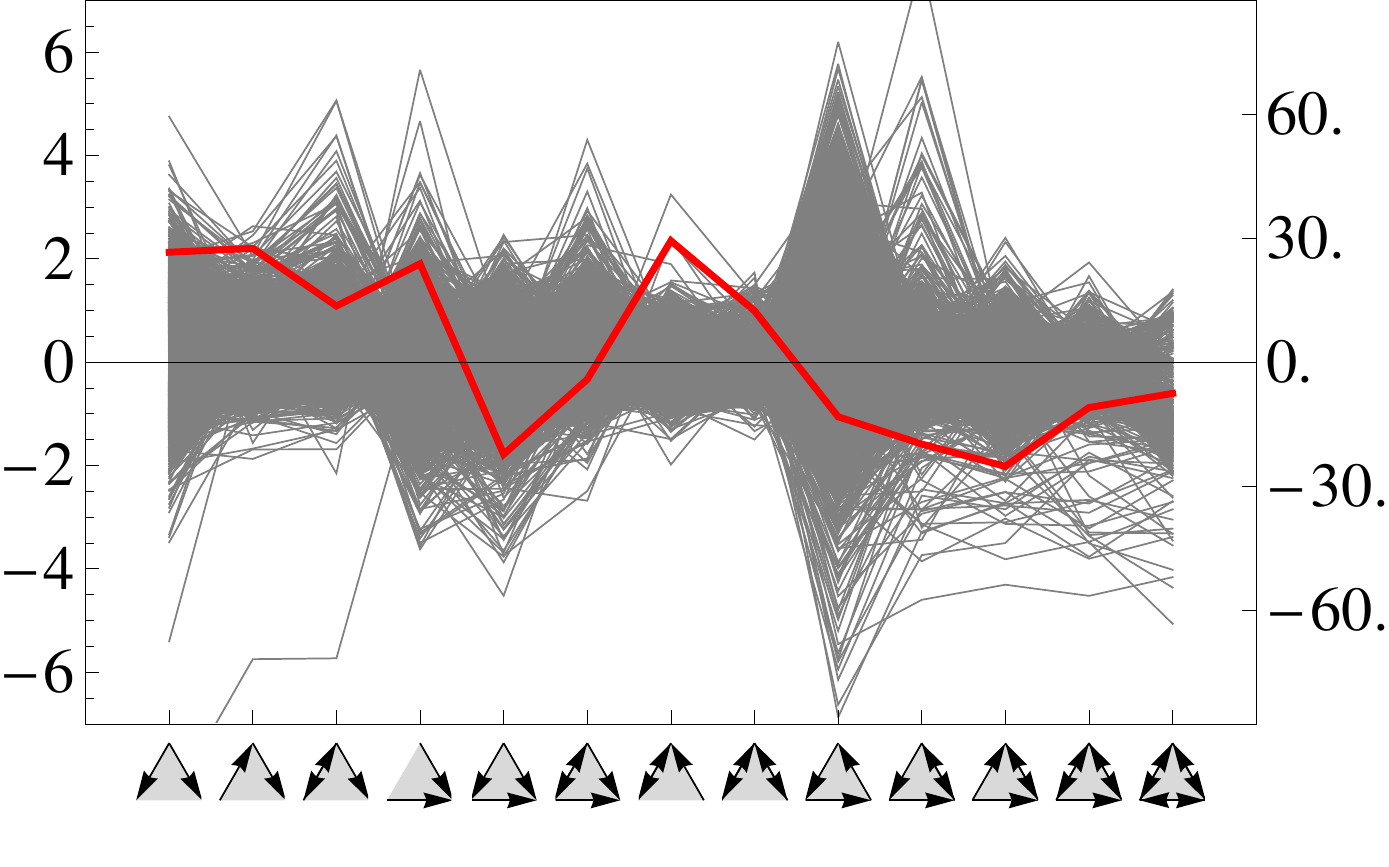}
		\label{fig:nspZ_profiles_mapped_mean_spanishBookInter}
	}
	\caption{Node-specific triadic Z-scores mapped to the patterns of Figure~\ref{fig:triadPatterns}. For each pattern the average is taken over all corresponding node-specific patterns (Table~\ref{tab:patternMapping}). The scaling on the left corresponds to the node-specific triad patterns, the one on the right to the Z-scores of the ordinary triad patterns.}
	\label{fig:nspZ_profiles_mapped_mean}
\end{figure*}

The gray, thin curves in Figure~\ref{fig:nspZ_profiles_mapped_mean} show the mapped scores for each node for multiple real-world networks. In addition, the red, thick curve shows the regular Z-score profile over the whole network obtained by the well-established motif-detection analysis of Milo et al. \cite{Milo2004}. Although the gray and the red curves are not independent of each other, it shall be mentioned that the regular Z-score profile can not be computed from the gray curves directly. In particular it is not the mean of the latters.

It can be observed that even though a pattern may be overrepresented referring to the system as a whole, it may still be underrepresented in the neighborhood of certain nodes. Moreover, there are patterns with a rather low regular Z-score, while there are both nodes with a strong positive and nodes with a strong negative contribution to the pattern. These contradictory effects seem to compensate each other on the system level. The decribed phenomenon can be particularly observed in the word-adjacency networks in Figures~\ref{fig:nspZ_profiles_mapped_mean_frenchBookInter} and~\ref{fig:nspZ_profiles_mapped_mean_spanishBookInter}, especially for the loop pattern ($\motXII$).

\subsection{Heterogeneity of motif distributions}

\begin{figure*}
	\centering
	\subfigure[][Yeast]{
		\includegraphics[width=0.36\textwidth]{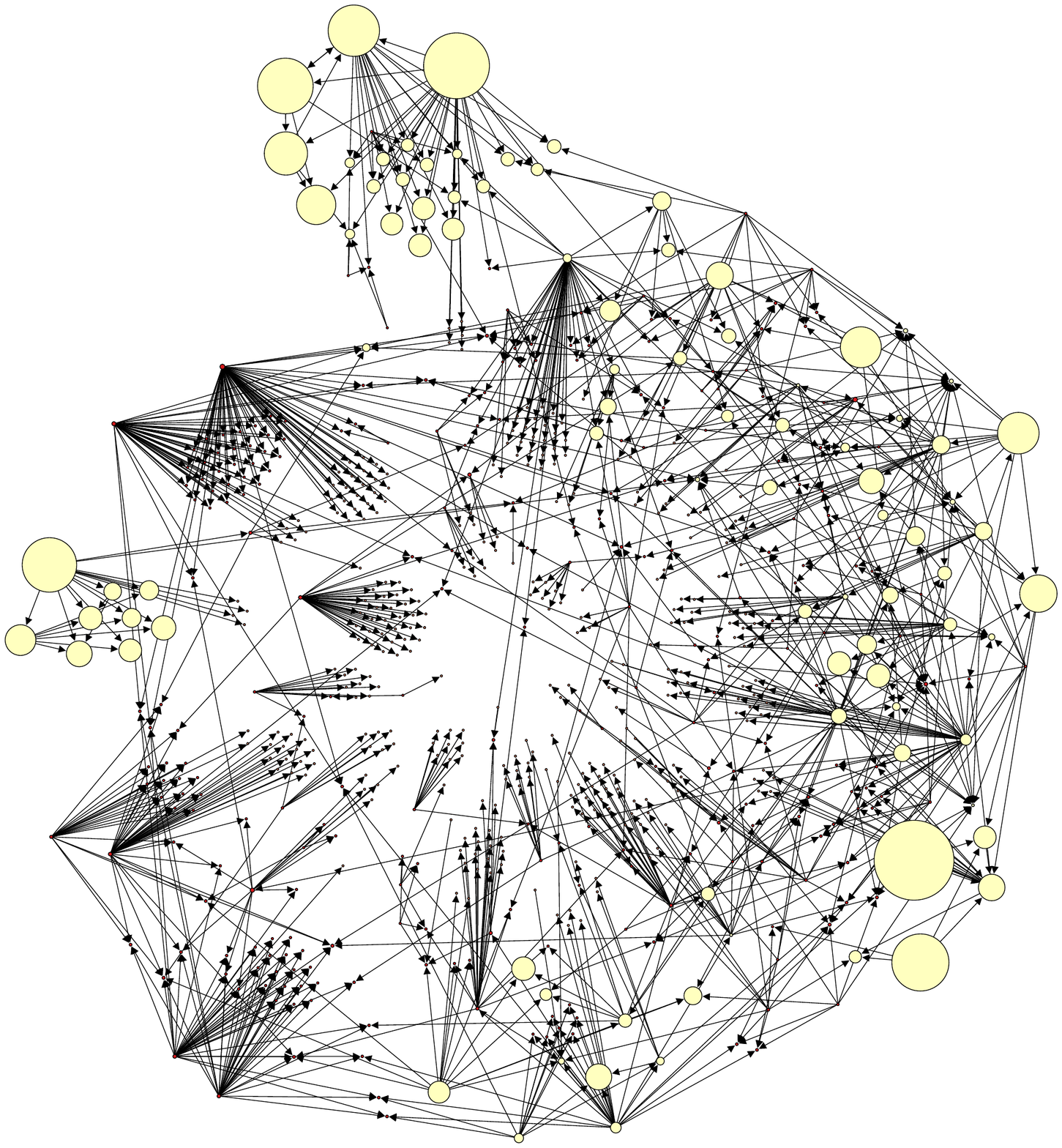}
		\label{fig:yeast_graphPlot_gephi_connected}
	}
	\subfigure[][E. Coli]{
		\includegraphics[width=0.34\textwidth]{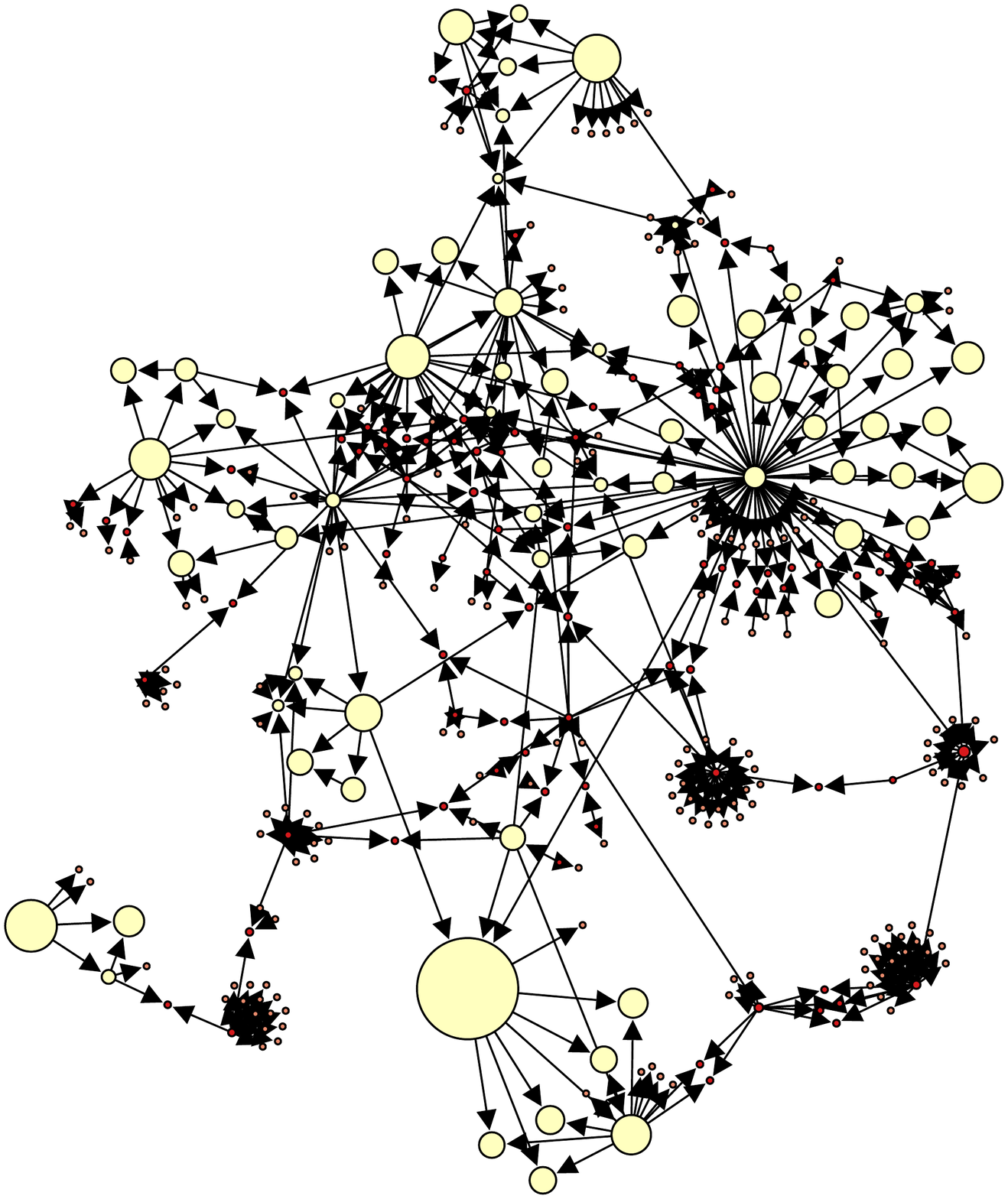}
		\label{fig:coli_graphPlot_gephi_connected}
	}
	\caption{Transcriptional regulatory networks. Vertex sizes indicate the magnitude of the mean node-specific triad Z-scores corresponding to the feed-forward loop motif (numbers 14, 16, and 23 in Figure~\ref{fig:dir_triadLocalPatterns_connected}). White vertices indicate positive values, filled vertices correspond to negative values (occur only with very small, hardly visible nodes). Graph plots were produced with Gephi \cite{gephi}.}
	\label{fig:graphPlots}
\end{figure*}

To further investigate whether motifs appear homogenously distributed over a graph we will devote ourselves to the feed-forward loop (FFL) pattern ($\motVIII$). The FFL is one of the patterns which has been studied most intensively with respect to its relevance for systems to reliably perform their functions \cite{Shen-Orr2002,Mangan2003,Mangan2003_2,Alon2007}. Specifically in transcriptional regulation networks it was argued that the FFL pattern might play an important role for facilitating its information-processing tasks~\cite{Shen-Orr2002}.

Figure~\ref{fig:graphPlots} shows two of those transcriptional regulation networks. In both of them the FFL is a motif. Vertex sizes are scaled by the magnitude of the averaged node-specific Z-scores of the three patterns corresponding to the FFL. Positive contributions are shown with bright vertices, negative ones are filled (do only occur with very small magnitude, i.e. very small node sizes). 
Apparently there are no nodes in the networks with a significant negative contribution to the FFL. Yet, neither is the pattern homogenously overrepresented throughout the whole system, although it is a motif. In fact, for most nodes the FFL-subgraph structure does not seems to play any role whatsoever. In contrast, there are few nodes with a rather strong contribution to the FFL eventually making it a motif of the entire system. 

\begin{figure*}
\newcommand{\w}{0.37}
	\centering
	\subfigure[][Yeast transcriptional]{
		\includegraphics[width=\w\textwidth]{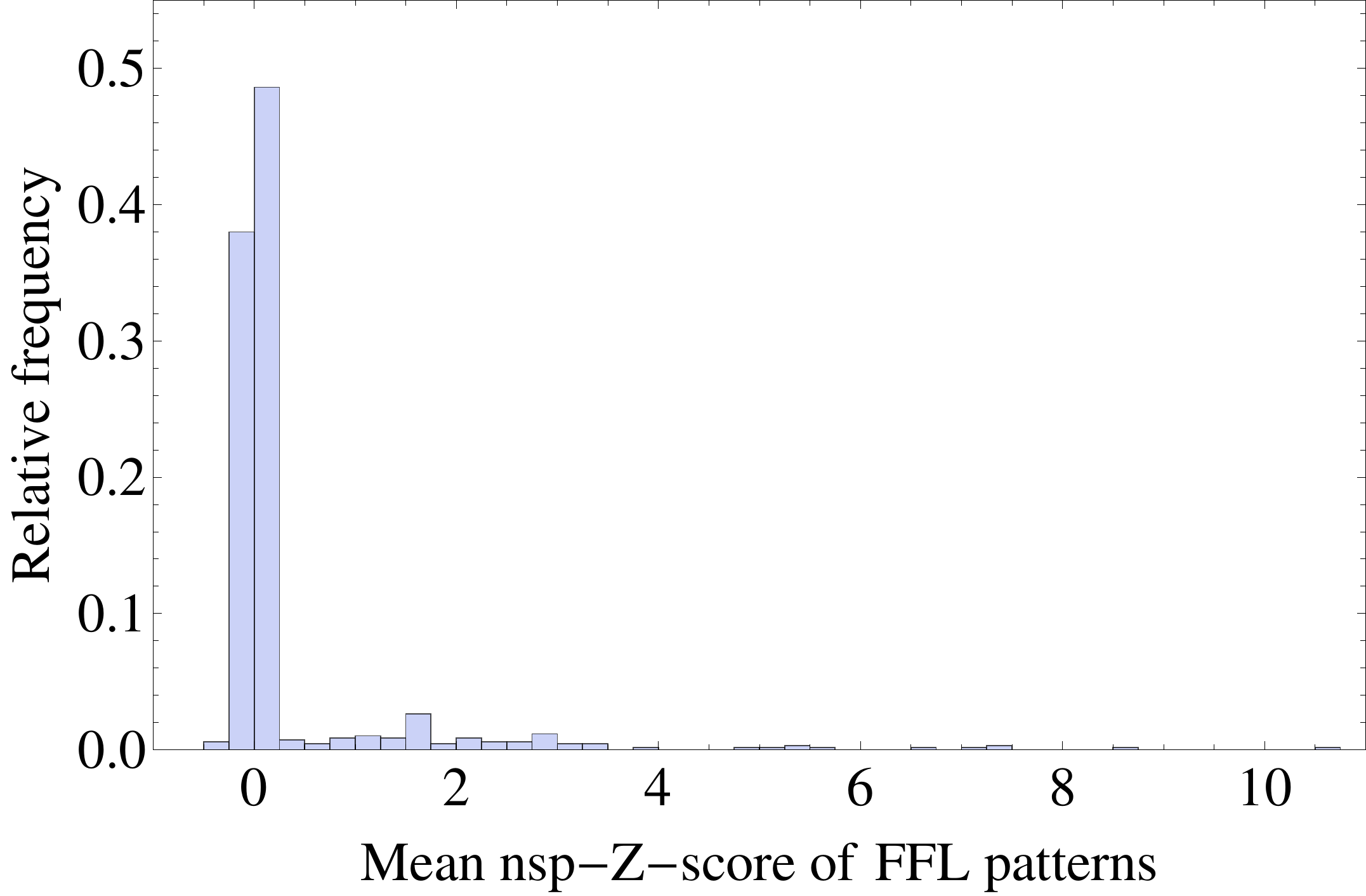}
		\label{fig:nspZ_profiles_histogramFeedForwardLoop_yeastInter}
	}
	\subfigure[][E. Coli transcriptional]{
		\includegraphics[width=\w\textwidth]{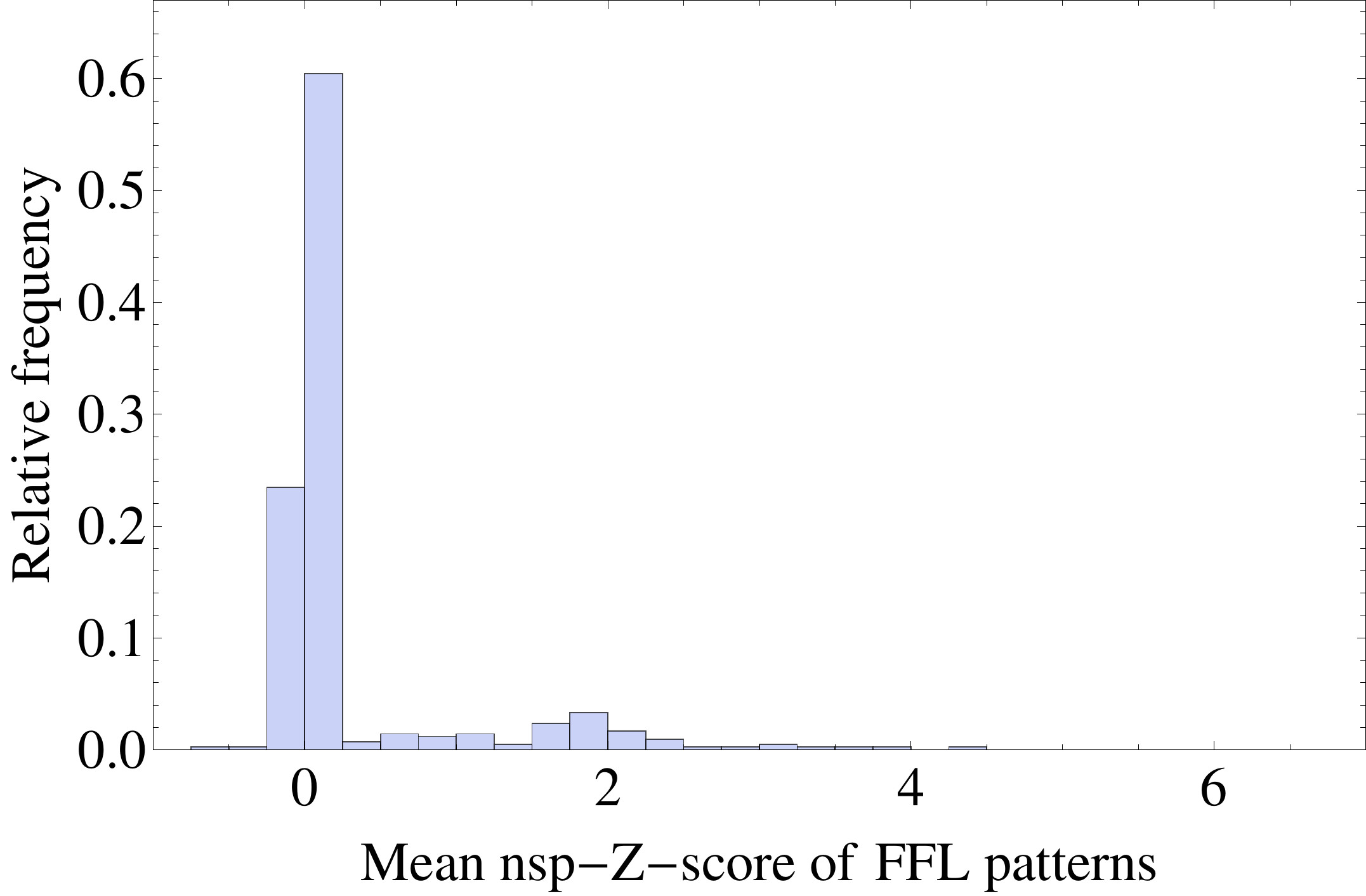}
		\label{fig:nspZ_profiles_histogramFeedForwardLoop_coli1_1Inter}
	}
	\caption{Histograms for the mean node-specific triad Z-scores corresponding to the feed-forward loop motif (numbers 14, 16, and 23 in Figure\ref{fig:dir_triadLocalPatterns_connected}).}
	\label{fig:nspZ_profiles_histogramFeedForwardLoop}
\end{figure*}

This effect becomes even clearer when considering histograms over the nodes' FFL contributions of the two systems. Figure~\ref{fig:nspZ_profiles_histogramFeedForwardLoop_yeastInter} shows the histogram of S. Cerevisiae, Figure~\ref{fig:nspZ_profiles_histogramFeedForwardLoop_coli1_1Inter} the one of E. Coli. Both exhibit a strong peak around zero indicating that most nodes do not participate in FFL structures significantly. Only very few nodes have a large mean node-specific triadic Z-score for the patterns corresponding to the FFL. 

There are two potential implications which can be derived from these observations: One could be that the FFL motif is actually not that important for the systems to work reliably. The second consequence could be that in fact very few nodes are critical for the system to work the way it is supposed to do. In the second case system would be very prone to failure of these crucial vertices. It may be subject to future research to further investigate these possible implications for dynamical processes on different topologies and under node failure.

\section{Conclusions} \label{sec:conclusions}

With this work we have introduced a novel tool for the analysis of complex networks in terms of their local substructure. Existing methods have focussed on the detection of patterns, which appear more frequently than expected for a random null model, over the network \textit{as a whole}. However, it has remained unclear whether these network motifs are overrepresented homogenously over the systems or whether they are bound to the neighborhood of certain nodes in the networks.

To overcome this limitation we introduced the framework of node-specific pattern mining, \textsc{NoSPaM}. Rather than averaging over the local topology of the entire system, for every node in the graph, we evaluate Z-score profiles which describe the nodes' individual local topologies.

For the analyis of real-world data sets we applied \textsc{NoSPaM}$_3$ which analyzes the local topology in terms of triadic subgraph patterns. We found that systems of similar fields tend to have similar node-specific triadic Z-score profiles indicating that local structural aspects are intimately connected with the systems' function.

Considering the mean contribution of node-specific triadic Z-scores to their respective ordinary triad patterns we found that the appearance of certain subgraphs is distributed highly heterougenously for many systems. This observation was supported by investigating the appearance of the feed-forward loop (FFL) pattern in more detail for transcriptional regulation networks in which it is a motif. The functional relevance of the FFL for systems to properly perform their task has been discussed intensively in complex networks research. It was conjectured that the FFL plays a key role for systems to process information. The fact that it appears highly heterogenously distributed over the graphs raises the question about its actual role for the systems. Will their function be significantly shortened when these nodes fail or are motifs in fact not important to such an extent? And how is the evolution of dynamical processes on systems affected by nodes with certain subgraph characteristics? To answer these questions will be subject to future research.

It may be further promising to adapt the analysis to signed networks in order to investigate issues related to social balance research. Moreover, the \textsc{NoSPaM} analysis of a network yields a new set of structural features for each node. Using these properties as inputs for clustering, classification, or role-detection algorithms can give rise to a better understanding of network designs and to improve link prediction on graphs.

\bibliographystyle{IEEEtran}
\bibliography{references}

\end{document}